\documentstyle[11pt,epic,eepic]{article}
\topmargin - 1.0 true cm
\newcommand{\beq}{\begin{equation}}
\newcommand{\eeq}{\end{equation}}
\def\bea{\begin{eqnarray}}
\def\eea{\end{eqnarray}}
\def\nn{\nonumber}
\def\sss{\scriptscriptstyle}
\def\dcp{D_{\sss CP}}
\def\bd{B_d^0}
\def\bdbar{{\overline{B_d^0}}}
\def\bs{B_s^0}
\def\bsbar{{\overline{B_s^0}}}
\def\barp{{\raise.35ex\hbox
{${\sss (}$}}---{\raise.35ex\hbox{${\sss )}$}}}
\def\bdbarp{\hbox{$B_d$\kern-1.4em\raise1.4ex\hbox{\barp}}}
\def\bsbarp{\hbox{$B_s$\kern-1.4em\raise1.4ex\hbox{\barp}}}
\def\ks{K_{\sss S}}
\def\roughly#1{\mathrel{\raise.3ex\hbox
{$#1$\kern-.75em\lower1ex\hbox{$\sim$}}}}
\def\lsim{\roughly<}
\def\gsim{\roughly>}
\def\ijmp#1#2#3{{\it Int.\ J.\ Mod.\ Phys.} {\bf A#1} (19#2) #3}
\def\mpla#1#2#3{{\it Mod.\ Phys.\ Lett.} {\bf A#1}, #3 (19#2)}
\def\npb#1#2#3{{\it Nucl.\ Phys.} {\bf B#1}, #3 (19#2)}
\def\plb#1#2#3{{\it Phys.\ Lett.} {\bf B#1}, #3 (19#2)}
\def\prd#1#2#3{{\it Phys.\ Rev.} {\bf D#1}, #3 (19#2)}
\def\prep#1#2#3{{\it Phys.\ Rep.} {\bf C#1}, #3 (19#2)}
\def\prl#1#2#3{{\it Phys.\ Rev.\ Lett.} {\bf #1}, #3 (19#2)}
\def\zpc#1#2#3{{\it Zeit.\ Phys.} {\bf C#1}, #3 (19#2)}
\def \stone{{\it B Decays}, edited by S. Stone 
(World Scientific, Singapore, 1994)}
\def\Bsll{B \rightarrow X_s l^+ l^-} 
\def\Bsga{B \rightarrow X_s \gamma}
\begin{document}
\baselineskip=7truemm
\begin{flushright}
{\bf hep-ph/9807254} \\
HUPD-9820 \\
SNUTP 98-070 \\
KEK-TH-578\\
\end{flushright}

\begin{center}
\bigskip

{\Large \bf A Model Independent Analysis of the Rare $B$ Decay $\Bsll$ }\\ 
\bigskip

S. Fukae$^{a,}$\footnote{fukae@ipc.hiroshima-u.ac.jp},~~
C. S. Kim$^{b,}$\footnote{kim@cskim.yonsei.ac.kr,~~
http://phya.yonsei.ac.kr/\~{}cskim/ },~~ 
T. Morozumi $^{a,}$\footnote{morozumi@theo.phys.sci.hiroshima-u.ac.jp}~~ 
and~~ T. Yoshikawa$^{c,}$\footnote{JSPS Research Fellow,~~
yosikawa@theory.kek.jp}
\end{center}


\begin{flushleft}
~~~~~~~~~~~~~$a$: {\it Department of Physics, Hiroshima University,
 Higashi Hiroshima 739-8526, Japan}
\\
~~~~~~~~~~~~~$b$: {\it Department of Physics, Yonsei University, 
Seoul 120-749, Korea}\\
~~~~~~~~~~~~~$c$: {\it Theory Group, KEK, Tsukuba, Ibaraki 305-0801, Japan}
\end{flushleft}

\begin{center}

\bigskip
(\today)

\bigskip 

{\bf Abstract}

\end{center}

\begin{quote}
The most general model-independent analysis of the rare $B$ decay, 
$\Bsll$, is presented.
There are ten independent local four-Fermi interactions
which may contribute to this process. The branching ratio,
the forward-backward asymmetry, and the double differential
rate are written as functions
of the Wilson coefficients of the ten operators.  We also study
the correlation between the branching ratio and the
forward-backward asymmetry by changing each coefficient. 
This procedure tells us which types of operator contribute to
the process, 
and it will be  very useful to pin down new physics systematically,
once we have the experimental data with high statistics
and the deviation from the Standard Model is found.
\end{quote}
\newpage

\section{Introduction}

Rare $B$-meson decays are very useful for constraining new physics beyond 
the Standard Model (SM). 
In particular, the processes $\Bsga$ and $\Bsll$ are 
experimentally clean, and are possibly the most
sensitive to the various extensions to the SM because these decays occur
only through loops in the SM. 
Nonstandard model effects can manifest themselves in these rare decays 
through the
Wilson coefficients, which can have values distinctly different from their
Standard Model counterparts. 
(See for example,
\cite{Goto,Wells,handoko,Wyler,Grossman,Rizzo,Jang,Cho}.)
Compared to $\Bsga$,  
the flavor changing leptonic decay $\Bsll$ is more sensitive to the actual
form of the new interactions since we can measure experimentally
various kinematical distributions as well as a total rate.
While new physics will change only the systematically uncertain normalization
for $\Bsga$, the interplay of various operators will also change the spectra
of the decay $\Bsll$. 

We note that the previous studies towards the model independent analysis
have been limited to the subset of the ten local four-Fermi
interactions within the specific extended models,  such as
the two-Higgs-doublet model, the minimal supersymmetric model, 
the left-right symmetric model, and {\it etc}. 
(See for example, \cite{Wyler,AGM,Hewett}.)
In the SM and in many of its extensions, the decay $\Bsll$ is 
completely determined phenomenologically 
by the numerical values of Wilson coefficients
of only three operators evaluated at the scale $\mu \sim m_b$.
However, the most interesting case would be
that this three parameter fit is found  
unsuccessful to explain the real experimental distributions, 
and that the new interactions necessarily involve
an extension of the full operator basis to include new operators 
beyond the usual set \cite{Rizzo}. 
And, therefore, the new physics senario can be much richer
than any of those models.

In the present paper, instead of doing  a
model by model analysis, we analyse  $\Bsll$
in the most general model-independent way. 
There are ten types of local four-Fermi
interactions which may contribute to the process.
This is contrasted to the SM in which  there
 are only two local interactions coming from the Box, 
$\gamma$ and  Z penguin diagrams and one non-local
interaction  coming from the cascade decay,
$B \rightarrow X_s \gamma^\ast \rightarrow X_s l^+ l^-$
\cite{InamiLim,Hou,Grinstein,Misiak,Buras}.
Our final purpose is to develop the most general way 
to distinguish the various four-Fermi
interactions, which can be fully investigated at $B$ factories of KEK-B, 
SLAC-B, B-TeV and LHC-B.
As is well known, all results in direct muon decay 
(energy spectra, polarizations
and angular distributions) and in inverse muon decay 
(the reaction cross section)
at energies well below $M_W$ may be parametrized \cite{Michel} in terms of
10 (complex) couplings, 
$g^\gamma_{\alpha\beta}~(\alpha,\beta=R,L,\gamma=S,V,T)$,
and the Fermi coupling constant $G_F$ using the matrix element,
\beq
{4 G_F \over \sqrt{2}} \sum_{\alpha,\beta,\gamma} 
g^\gamma_{\alpha\beta} <\bar e_\alpha | \Gamma^\gamma | \nu_e >
<\bar \nu_\mu | \Gamma_\gamma | \mu_\beta > .
\eeq
The ``$V-A$ interaction'' of the SM corresponds 
to the single coupling $g^V_{LL}$,
which has been experimentally probed thoroughly.
 
As quantities to be measured in our first step, 
we study the decay rate $\cal{B}$
and the forward-backward (FB) asymmetry ${\cal A}$ \cite{ATM}.
These are written as
functions of the coefficients of the interactions, {\it i.e.}, 
${\cal B}(\{C_i\})$ and ${\cal A}(\{C_i\})$, 
where $\{C_i\}$ is a set of
Wilson coefficients. We study the branching ratio and the FB
asymmetry by changing each coefficient.
This shows how each operator contributes to the
measurements in a different way from the other operators. 
Furthermore, to distinguish
types of operators, we draw the correlation 
between the decay rate and the asymmetry.
 By changing each coefficient, we can draw a
flow in the two dimensional plane $({\cal B}(C_m),{\cal A}(C_m))$ \cite{Okada}.
This flow depends on the type of the interaction which acts.
Thus we may see which type of operator contributes to the
process once these quantities are measured precisely. 

The paper is organized as follows.
In Section 2 we show the most general four-Fermi
interactions and compute the decay distribution.
We also derive the branching ratio and the
asymmetry. Their dependence on the four-Fermi interactions
is also studied. In Section 3 the correlation between the
branching ratio and the asymmetry is discussed and Section
4 is devoted to our summary.  

\section{The four-Fermi interactions and the decay
distributions}

We start by defining the various kinematic variables.
In the present paper, the inclusive semileptonic
$B$ decay is modeled by the partonic calculation, {\it i.e.},
$b(p_b) \rightarrow s(p_s) + l^+(p_+) + l^-(p_-) $. 
This is regarded as the leading order calculation in the $1/m_b$
expansion \cite{Falk,AGHM}.  
Then the decay distribution is described  
by the following two kinematic variables $s$ and $u$, 
\bea
s &=& ( p_b - p_s )^2 = ( p_+ + p_- )^2  
         = m_b^2 + m_s^2 + m_+^2 + m_-^2 - t_+ - t_- ,\nn \\
t_+ &=& ( p_s + p_+ )^2 = ( p_b - p_- )^2, \nn \\
t_- &=& ( p_s + p_- )^2 = ( p_b - p_+ )^2 ,\nn \\
u &=& t_+ - t_- .
\eea
In the center of mass frame of the dileptons,
$u$ is written in terms of 
$\theta $, {\it i.e.}, the angle between the momentum of the $B$ meson
and that of $l^+$,
\bea
u &=& - u(s) \cdot cos\theta \equiv - u(s) z,\nn \\  
{\rm with}~~~~~~z &=& \cos\theta,\nn \\
u(s) &=& \sqrt{(s-(m_b+m_s)^2)(s-(m_b-m_s)^2)(1 -
                            \frac{4m_l^2}{s})}.
\eea
The phase space is defined in terms of  $s$ and $z$, 
\bea
4 m_l^2 \leq s  \leq ( m_b - m_s)^2, \nn \\
-1 \leq z \leq 1.
\eea

The matrix element of the decay is written as the sum of the
SM contribution and the contribution from the
local four-Fermi interactions,
\beq
{\cal M} = {{\cal M}_{_{\rm SM}}} + {{\cal M}_{_{\rm NEW}}},
\eeq
where ${{\cal M}_{_{\rm SM}}}$ is the SM part and is given by
\bea
{\cal M}_{_{\rm SM}} = \frac{G_F \alpha }{\sqrt{2}\pi }V_{ts}^*V_{tb}
              &[& ( C_9^{eff} - C_{10})~\bar{s}_L \gamma_\mu b_L 
                           ~\bar{l}_L \gamma^\mu l_L  \nn \\
              &+&  ( C_9^{eff} + C_{10})~\bar{s}_L \gamma_\mu b_L  
                           ~\bar{l}_R \gamma^\mu l_R  \nn \\
              &-& 2 C_7^{eff}  ~\bar{s} i \sigma_{\mu \nu } \frac{q^\nu}{q^2} 
                        ( m_s L + m_b R ) b ~\bar{l} \gamma^\mu l ~], 
\eea
We use the following values for the Wilson
coefficients $ ( C_9^{NDR}= 4.153, C_{10}=-4.546, C_7= - 0.311 )$.
They correspond to the next-to leading calculation of
QCD corrections \cite{Misiak,Buras}.
The renormalization point $\mu$ and the top quark mass are  set
to be $\mu = m_b=4.8~({\rm GeV})$ and $m_t=175~({\rm GeV})$,
unless otherwise specified. (We follow Ref. \cite{KMS} 
for the choice of the parameters in the SM as well as the 
incorporation of the long-distance effects of charmonium states 
\cite{Long,Kruger}.)
There are ten independent local four-Fermi interactions which may
contribute to the process.  ${\cal M}_{_{\rm NEW}}$ is a function
of the coefficients of local four-Fermi interactions and
is defined as 
\bea
{\cal M}_{_{\rm NEW}} = \frac{G_F \alpha }{\sqrt{2}\pi }V_{ts}^*V_{tb} 
              &[& C_{LL} ~\bar{s}_L \gamma_\mu b_L 
                           ~\bar{l}_L \gamma^\mu l_L  \nn \\
              &+& C_{LR} ~\bar{s}_L \gamma_\mu b_L  
                           ~\bar{l}_R \gamma^\mu l_R  \nn \\
              &+& C_{RL} ~\bar{s}_R \gamma_\mu b_R 
                           ~\bar{l}_L \gamma^\mu l_L  \nn \\
              &+& C_{RR} ~\bar{s}_R \gamma_\mu b_R  
                           ~\bar{l}_R \gamma^\mu l_R  \nn \\
              &+& C_{LRLR} ~\bar{s}_L b_R ~\bar{l}_L l_R \nn \\
              &+& C_{RLLR} ~\bar{s}_R b_L ~\bar{l}_L l_R \nn \\
              &+& C_{LRRL} ~\bar{s}_L b_R ~\bar{l}_R l_L \nn \\  
              &+& C_{RLRL} ~\bar{s}_R b_L ~\bar{l}_R l_L \nn \\
              &+& C_T       ~\bar{s} \sigma_{\mu \nu } b 
                           ~\bar{l} \sigma^{\mu \nu } l \nn \\
              &+& i C_{TE}   ~\bar{s} \sigma_{\mu \nu } b 
                           ~\bar{l} \sigma_{\alpha \beta } l 
                           ~\epsilon^{\mu \nu \alpha \beta } ] ,
\eea
where the $C_{X}$'s are the coefficients of
four-Fermi interactions. 
Among them, there are four vector type interactions denoted by
$C_{LL}$, $C_{LR}$,   $C_{RL}$, and  $C_{RR}$.
Two of them ($C_{LL}$, $C_{LR}$) are already present in the
SM as the combinations of  ($C_9-C_{10}$, $C_9+C_{10}$).
Therefore they correspond to the deviation of these coefficients
from the SM values.
The other interactions, denoted by  $C_{RL}$ and $C_{RR}$,
are obtained by interchanging  the chirality projections
$L$ and $ R $.
There are four scalar type interactions, $C_{LRLR}$, $C_{RLLR}$,
$C_{RLLR}$, and  $C_{RLRL}$. The remaining two, denoted by $C_T$
and  $C_{TE}$, correspond to tensor type.   

\def\BXsee{B \rightarrow X_s e^+e^-}
\def\BXsmumu{B \rightarrow X_s \mu^+\mu^-}

It is now straightforward to compute the double differential
decay rate. The results for the massless lepton case, 
$\BXsee$ and $\BXsmumu$,
are explicitly written in the text. The massive
case is separately given in the Appendix.
The double differential rate is a function of
the dilepton invariant mass $s$ and $z=\cos\theta$.
In the massless limit, the interference between 
tensor (scalar) type interactions  and vector type interactions
vanishes. Then the expression becomes considerably
simplified, 
\bea
\frac{d {\cal B}}{d s d z} = \frac{1}{2 {m_b}^8} {\cal B}_0 &[& 
                 A_1(s,z)~\{ 4 \left|C_7\right|^2 \} \nn \\
             &+& A_2(s,z)~\{ \left|(C_9^{eff} - C_{10})\right|^2 
                         +  \left|(C_9^{eff} + C_{10})\right|^2 \} \nn \\
             &+& A_3(s,z)~\{ \left|(C_9^{eff} - C_{10})\right|^2 
                         -  \left|(C_9^{eff} + C_{10})\right|^2 \} \nn \\
             &+& A_4(s,z)~\{ 2 Re(-2 C_7 ( C_9^{eff *} - C_{10}^* )) 
                          + 2 Re(-2 C_7 ( C_9^{eff *} + C_{10}^* )) \} \nn \\
             &+& A_5(s,z)~\{ 2 Re(-2 C_7 ( C_9^{eff *} - C_{10}^* )) 
                          - 2 Re(-2 C_7 ( C_9^{eff *} + C_{10}^* )) 
                                                             \} \nn \\[5mm]
             &+& A_2(s,z)~\{ 2 Re(( C_9^{eff} - C_{10}){\bf C_{LL}^* })  
                          + 2 Re(( C_9^{eff} + C_{10}){\bf C_{LR}^* }) 
                                                            \} \nn \\
             &+& A_3(s,z)~\{ 2 Re(( C_9^{eff} - C_{10}){\bf C_{LL}^* })  
                          - 2 Re(( C_9^{eff} + C_{10}){\bf C_{LR}^* })
                                                            \} \nn \\
             &+& A_4(s,z)~\{ 2 Re( -2 C_7({\bf C_{LL}^* + C_{LR}^* }))\} \nn \\
             &+& A_5(s,z)~\{ 2  Re( -2 C_7({\bf C_{LL}^* - C_{LR}^* }))
                                                            \} \nn \\ 
             &+& A_6(s,z)~\{ 2 Re( -2 C_7({\bf C_{RL}^* + C_{RR}^* }))\}
                                                               \nn \\
             &+& A_6(s,z)~\{ - 2 Re(( C_9^{eff} - C_{10}){\bf C_{RL}^* })  
                            - 2 Re(( C_9^{eff} + C_{10}){\bf C_{RR}^* })
                                                            \} \nn \\
             &+& A_7(s,z)~\{ 2 Re( -2 C_7({\bf C_{RL}^* - C_{RR}^* })) 
                                                         \} \nn \\[5mm]
             &+& A_2(s,z)~\{ {\bf \left|C_{LL}\right|^2 + \left|C_{LR}\right|^2 
             + \left|C_{RL}\right|^2 + \left|C_{RR}\right|^2  } \} \nn \\
             &+& A_3(s,z)~\{ {\bf  \left|C_{LL}\right|^2 
                                - \left|C_{LR}\right|^2 
                                -  \left|C_{RL}\right|^2 
                                + \left|C_{RR}\right|^2 } \} \nn \\
             &+& A_3(s,z)~\{ - 4 {\bf Re[ C_{LRLR} (C_{T}^* - 2 C_{TE}^* )
                                  + C_{RLRL} (C_{T}^* + 2 C_{TE}^* )]} 
                                                               \} \nn \\
             &+& A_6(s,z)~\{ - 2 {\bf Re( C_{LL} C_{RL}^* 
                                      + C_{LR} C_{RR}^* )} \nn \\
             & & ~~~~~~~~~ + {\bf Re(C_{LRLR} C_{RLLR}^* 
                            + C_{LRRL} C_{RLRL}^* ) } \} \nn \\
             &+& A_8(s,z)~\{ {\bf \left|C_{LRLR}\right|^2 
                                + \left|C_{RLLR}\right|^2 
                                + \left|C_{LRRL}\right|^2 
                                + \left|C_{RLRL}\right|^2  } \} \nn \\
             &+& A_9(s,z)~\{ {\bf 16 \left|C_{T}\right|^2 
                                + 64 \left|C_{TE}\right|^2 } \} ], 
\eea
where $A_n$ are the functions of kinematic variables $s$ and $z$, 
\bea
A_1(s,z) &=& \frac{1}{s}u(s)\{ -16 m_b^2 m_s^2 s 
             + ( m_b^2 + m_s^2 )(- 2 s^2 + 2 u(s)^2 z^2 
             + 2 ( m_b^2 - m_s^2 )^2 ) \},  \\
A_2(s,z) &=& u(s)\{- u(s)^2 z^2 - s^2 + ( m_b^2 - m_s^2 )^2 \},  \\
A_3(s,z) &=&  2 u(s)^2 z s , \\
A_4(s,z) &=& u(s)\{ 2 ( m_b^2 + m_s^2 ) s - 2 ( m_b^2 - m_s^2 )^2 \},  \\
A_5(s,z) &=& - 2 ( m_b^2 + m_s^2 ) u(s)^2 z , \\
A_6(s,z) &=& 4 m_b m_s s u(s),  \\
A_7(s,z) &=& 4 m_b m_s u(s)^2 z , \\
A_8(s,z) &=& u(s)( m_b^2 + m_s^2 -s ) s , \\
A_9(s,z) &=& u(s)\{ - 2 u(s)^2 z^2 - 2 s (m_b^2 + m_s^2 ) 
                              + 2 ( m_b^2 - m_s^2 )^2  \},  \\
u(s) &=& \sqrt{ \{ s - ( m_b - m_s )^2 \}\{s - ( m_b + m_s )^2 \} }. 
\eea
Among them, $A_1,A_2, A_4, A_6, A_8, A_9$ are even
functions with respect to $z$ while $A_3,A_5$,and $A_7$
are odd functions with respect to z.
The former contribute to the single differential rate 
$d{\cal B}/ds$ and the latter contributes to the forward-backward (FB) 
asymmetry $d{\cal A}/ds$.

Let us first investigate the single differential rate,
\beq
\frac{d {\cal B}}{d s} = \int^1_{-1} \frac{d^2 {\cal B}}{d s
d z} dz. 
\eeq
It is given by
\bea
\frac{d {\cal B}}{d s}(s) = \frac{1}{2 {m_b}^8}{\cal B}_0 
            &[& M_1(s)~\{ 4 \left|C_7\right|^2 \} \nn \\
            &+& M_2(s)~\{ 2 \left| C_9^{eff} \right|^2 
                       + 2 \left|C_{10}\right|^2  \nn \\
            & & ~~~~~~~+ 2 Re( C_9^{eff} ( C_{LL}^* + C_{LR}^* ) )
                       + 2 Re( C_{10} ( - C_{LL}^* + C_{LR}^* )) \nn \\
         & & ~~~~~~~+ \left|C_{LL}\right|^2 + \left|C_{LR}\right|^2 
               + \left|C_{RL}\right|^2 + \left|C_{RR}\right|^2 \} \nn \\
       &-& M_4(s)~\{ 4 Re[C_7^* ( 2 C_9^{eff} + C_{LL} + C_{LR})] \} \nn \\
            &-& M_6(s)~\{ 4 Re[C_7^* (C_{RL} + C_{RR})] \nn \\
            & & ~~~~~~~+ 2 Re( C_9^{eff} ( C_{RL}^* + C_{RR}^* ) )
                       + 2 Re( C_{10} ( - C_{RL}^* + C_{RR}^* )) \nn \\
            & & ~~~~~~~+ 2 Re( C_{LL} C_{RL}^* + C_{LR} C_{RR}^* )
                       -   Re( C_{LRLR} C_{RLLR}^* 
                             + C_{LRRL} C_{RLRL}^* )\} \nn \\
            &+& M_8(s)~\{ \left|C_{LRLR}\right|^2 
                       + \left|C_{RLLR}\right|^2 
                       + \left|C_{LRRL}\right|^2 
                       + \left|C_{RLRL}\right|^2 \} \nn \\
            &+& M_9(s)~\{ 16 \left|C_{T}\right|^2 
                     + 64 \left|C_{TE}\right|^2 \} ], 
\eea
where
${\cal B}_0$ is a normalization factor,
\bea
{\cal B}_0 = {\cal B}_{sl} \frac{3 \alpha^2 }{16 \pi^2 }
             \frac{ |V_{ts}^*V_{tb}|^2 }{|V_{cb}|^2 } 
            \frac{1}{f(\hat{m_c}) \kappa(\hat{m_c})} ,
\eea
and the phase space factor, $f(\hat{m_c}={m_c \over m_b})$, 
and the $O(\alpha_s)$
QCD correction factor \cite{Kim}, $\kappa(\hat{m_c})$, 
of $b \rightarrow c l \nu$ are given by
\bea
f(\hat{m_c}) &=& 1 - 8 \hat{m_c}^2 + 8 
        \hat{m_c}^6 - \hat{m_c}^8 - 24 \hat{m_c}^4 \ln \hat{m_c} ,  \\
\kappa(\hat{m_c}) &=& 1 - \frac{2 \alpha_s(m_b)}{3 \pi} 
           [(\pi^2-\frac{31}{4})(1-\hat{m_c})^2 + \frac{3}{2} ] . 
\eea
For the numerical calculations, we set  
$\frac{ |V_{ts}^*V_{tb}|^2 }{|V_{cb}|^2 } =1$
and use  ${\cal B}_{sl}=10.4 \%$.
$M_n$  are  functions which are obtained
by integrating $A_n(s,z)$ with respect to $z$ as follows,
\bea
M_n(s) &=& \int_{-1}^1 A_n(s,z) dz = 2 A_n(s,\sqrt{1/3}),
\eea 
where $n$ is 1,2,4,6,8 and 9 and they are explicitly given by
\bea
M_1(s) &=& 2 u(s) \frac{1}{s} [ - 16 m_b^2 m_s^2 s + 
          ( m_b^2 + m_s^2 )( - 2 s^2 + \frac{2}{3}u(s)^2 
        + 2 ( m_b^2 - m_s^2 )^2 ) ],  \\
M_2(s) &=& - 2 u(s) [ \frac{1}{3}u(s)^2 + s^2 
          - ( m_b^2 - m_s^2 )^2 ],  \\ 
M_4(s) &=& 2 u(s) [ 2 ( m_b^2 + m_s^2 ) s - 2 ( m_b^2 - m_s^2 )^2 ],  \\
M_6(s) &=& 2 u(s) [ 4 m_b m_s s ],  \\
M_8(s) &=& 2 u(s) [ s ( m_b^2 + m_s^2 - s )],  \\
M_9(s) &=& 2 u(s) [ - \frac{2}{3}u(s)^2 - 2 ( m_b^2 + m_s^2 ) s 
                   + 2  ( m_b^2 - m_s^2 )^2 ].
\eea

In order to compare the kinematic functions $M_n$,
we plot them in Figure 1.
This plot shows that $M_1$ is the largest due to its $1/s$ dependence
and that $M_6$ is suppressed by a factor of $m_s$ (the strange quark mass).
Knowing the kinematical dependence, we plot the differential
decay rate $d{\cal B}/ds$ by changing each coefficient $C_{X}$
in the new physics amplitude ${{\cal M}_{_{\rm NEW}}}$.
 We summarize how the differential rate  
$d{\cal B}/ds$ is changed accordingly.
\begin{itemize}
\item{{\bf Vector type interactions}\\
We assume that $ C_X $ are real parameters. 
This means we do not introduce any new physics phase 
 in addition to 
 that of the SM given by Arg$(V_{ts}^\ast  V_{tb})$.
In Figs. 2,3,4 and 5, we change $C_{LL}, C_{LR}, C_{RL},
C_{RR}$ respectively.
Among the vector interactions, $d{\cal B}/ds$  changes the most 
with respect to $C_{LL}$. This can be explained as follows. 
By neglecting the term proportional to $M_6$,
we can collect the terms coming from $C_{LL}, C_{LR}, C_{RL},
$ and $C_{RR}$. They are given by 
\bea
M_2(s) C_{LL}^2 + \{ M_2(s) 2 ( C_9^{eff} - C_{10} )
                  + M_4(s) ( -4 C_7 ) \} C_{LL},   \\
M_2(s) C_{LR}^2 + \{ M_2(s) 2 ( C_9^{eff} + C_{10} )
                  + M_4(s) ( -4 C_7 ) \} C_{LR},  \\
M_2(s) C_{RL}^2 ,  \\
M_2(s) C_{RR}^2 . 
\eea 
Around $s = 5$ GeV$^2$, $Re( C_9^{eff} - C_{10} ) \sim 9.5$ and 
$Re( C_9^{eff} + C_{10} ) \sim 0.4$. Then the interference
term between $C_{LL}$ and $Re( C_9^{eff} - C_{10} )$
is large and  $d{\cal B}/ds$ is sensitive to the change of
$C_{LL}$. The contribution is constructive for $C_{LL}=|C_{10}|$
and destructive for $ C_{LL}=-|C_{10}|$, as shown in Figure 1.
The  effects of the other vector interactions
($C_{LR},C_{RR},C_{RL}$) are smaller than that of $C_{LL}$.
To see this dependence clearly,
we have also studied the
partially integrated branching ratio (${\cal B}$) and have shown
how it depends on the coefficients. (See Figures 8 and 9.)
The range for the integration is $1 \le s \le 8~ ({\rm GeV}^2).$
This range is below the $J/\psi$ resonance and $d{\cal B}/ds$ is nearly 
flat within the range.  From these figures, 
the branchingratio 
depends strongly on $C_{LL}$ while the dependence on the other
coefficients is rather weak. The contribution of $C_{LL}$ to
 ${\cal B}$ is positive (negative) for $C_{LL} >0 \quad (<0)$.
The contribution due to $C_{RL}$ and $C_{RR}$ 
is positive. }
\item{{\bf Scalar and tensor  type interactions}\\
The scalar and tensor interactions
increase $d{\cal B}/ds$, as shown in  Figures 6 and 7.
This can be interpreted as follows.
As far as we see the sensitivity to only one of
the scalar operators, {\it i.e.},
$C_{LRLR}, C_{RLLR}$, 
$ C_{LRRL}$ and $ C_{RLRL}$, the interference term does not act at all.
And it is enough to
study the response of $d{\cal B}/ds$ to the change of the
combinations of the coefficients,
$|C_{LRLR}|^2+ |C_{RLLR}|^2 + |C_{LRRL}|^2+ |C_{RLRL}|^2$
and   $16 |C_T|^2 + 64 |C_{TE}|^2$.
Because their contribution is always positive,
it increases $d{\cal B}/ds$. 
Furthermore, even if  two or more scalar interactions
act together,  the effect of the interference term denoted by $M_6$
is suppressed by a factor of $m_s$ 
and can not exceed the positive contribution.
Thus the major dependence on the coefficients
comes through the combination, 
$|C_{LRLR}|^2+ |C_{RLLR}|^2 + |C_{LRRL}|^2+ |C_{RLRL}|^2$,
even if two or more operators contribute to the process. 
Therefore, $d{\cal B}/ds$  increases even if more than one scalar
and/or tensor type of interaction are present.
In Figures 8 and 9, we can see that
the branching ratio  ${\cal B}$ always 
increases as one of the scalar and tensor interactions
act.
}
\end{itemize}

\bigskip 
\bigskip 
\bigskip 
\bigskip 
\bigskip 

\setlength{\unitlength}{0.240900pt}
\\
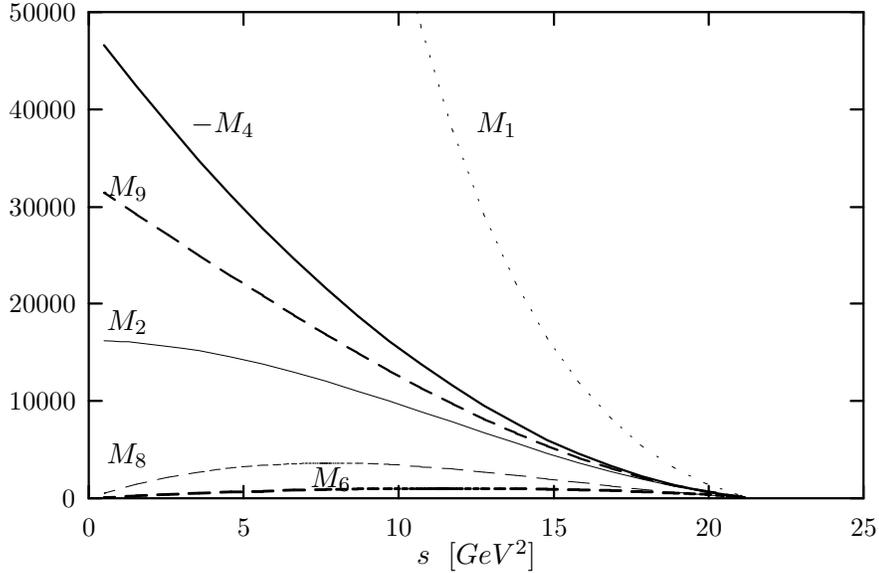
{Figure 1 ~~\small $M_1(s)$ (thin dotted line), $M_2(s)$ (thin solid
line), $- M_4(s)$ (thick solid line), $M_6(s)$ (thick
dashed line), $M_8(s)$ (thin dashed line) and $M_9(s)$ (dashed line).} 

\bigskip 
\bigskip 
\bigskip 
\bigskip

\setlength{\unitlength}{0.240900pt}
%
 \\
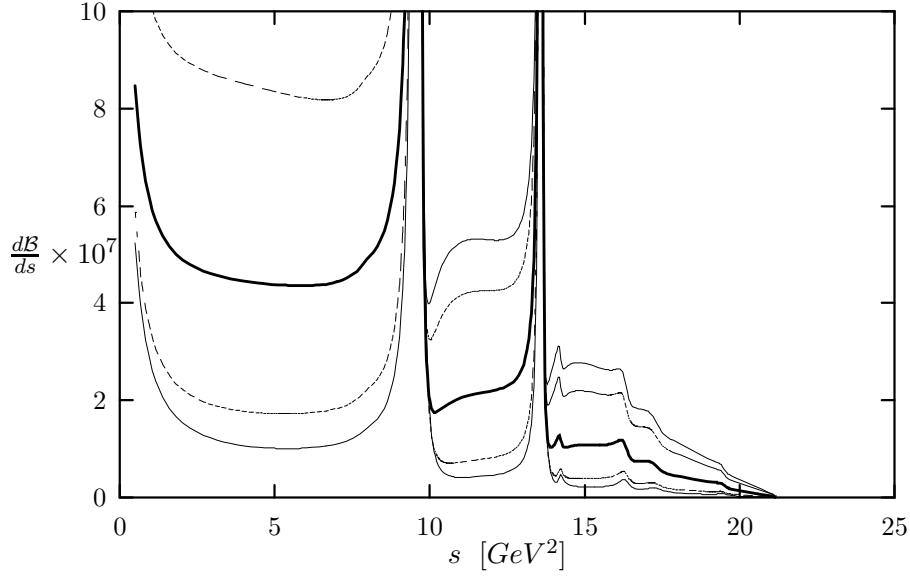
{Figure 2 ~~\small The differential branching ratio for the SM
(thick solid line), $ C_{LL} $ is $ - \left|C_{10}\right|$
(thin line), $ - 0.7 \left|C_{10}\right|$ (thin dashed
line), $ 0.7 \left|C_{10}\right|$ (thin dashed line) and
$ \left|C_{10}\right|$ (thin line). The coefficients of
the other interactions in ${\cal M}_{\rm NEW}$ are set to 
zero. } 
\bigskip 
\bigskip 
\bigskip 

\setlength{\unitlength}{0.240900pt}
%
 \\
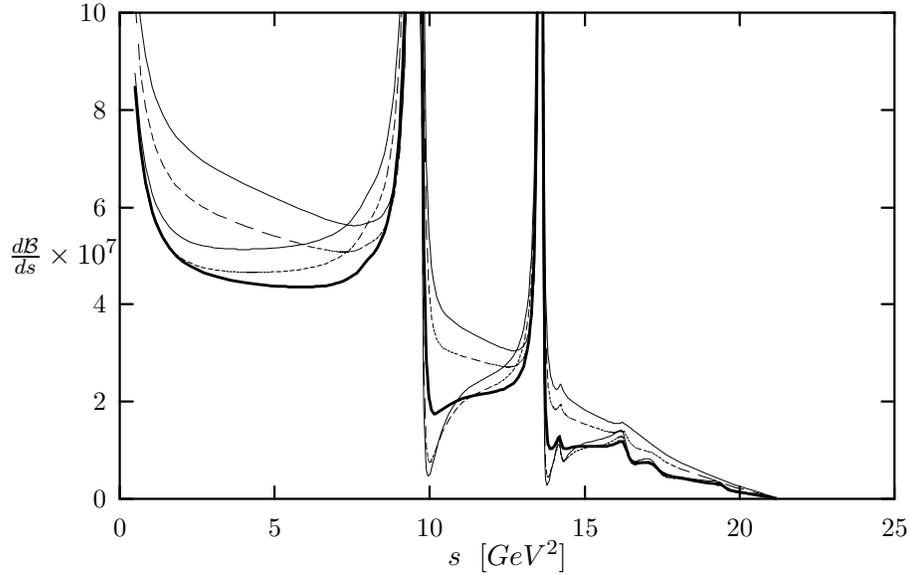
{Figure 3 ~~ \small The differential branching ratio for the SM
(thick solid line), $ C_{LR} $ is $ - \left|C_{10}\right|$
(thin line), $ - 0.7 \left|C_{10}\right|$ (thin dashed
line), $ 0.7 \left|C_{10}\right|$ (thin dashed line) and
$ \left|C_{10}\right|$ (thin line). The coefficients of
the other interactions in ${\cal M}_{\rm NEW}$ are set to zero.} 
\bigskip
\bigskip
\bigskip

\setlength{\unitlength}{0.240900pt}
%
 \\
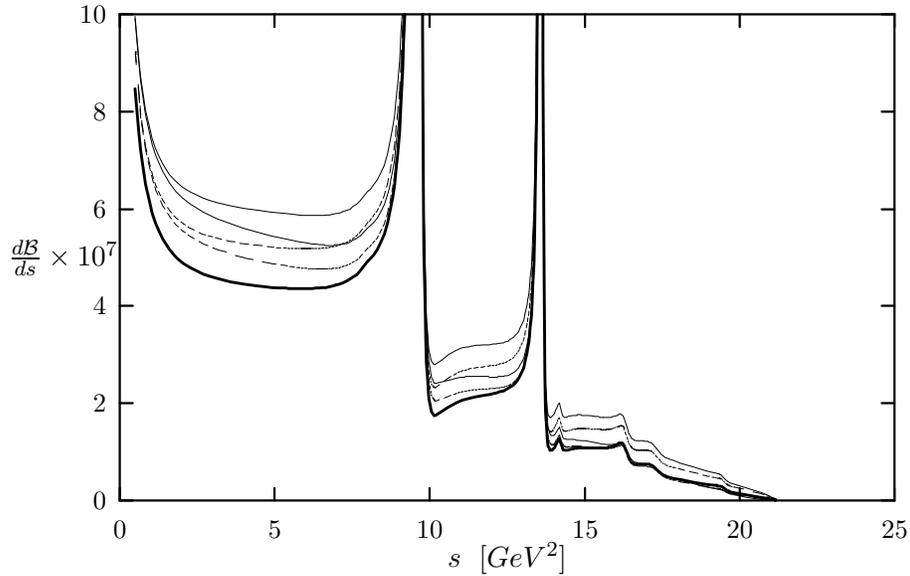
{Figure 4 ~~\small The differential branching ratio for the SM
(thick solid line), $ C_{RL} $ is $ - \left|C_{10}\right|$
(thin line), $ - 0.7 \left|C_{10}\right|$ (thin dashed
line), $ 0.7  \left|C_{10}\right|$ (thin dashed line) and
$ \left|C_{10}\right|$ (thin line).  The coefficients of
the other interactions in ${\cal M}_{\rm NEW}$ are set to zero.} 

\bigskip
\bigskip
\bigskip

\setlength{\unitlength}{0.240900pt}
%
 \\
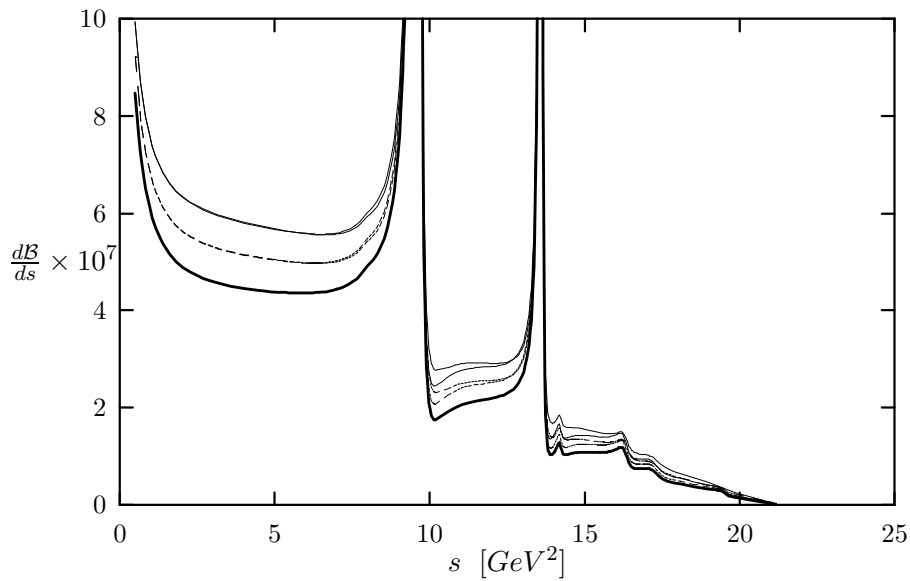
{Figure 5 ~~\small The differential branching ratio for the SM (thick
solid line), $ C_{RR} $ is $ - \left|C_{10}\right|$
(thin line), $ - 0.7 \left|C_{10}\right|$ (thin dashed
line), $ 0.7 \left|C_{10}\right|$ (thin dashed line) and
$ \left|C_{10}\right|$ (thin line). The coefficients of
the other interactions in ${\cal M}_{\rm NEW}$ are set to zero.  } 
\bigskip
\bigskip
\bigskip
\bigskip

\setlength{\unitlength}{0.240900pt}
%
 \\
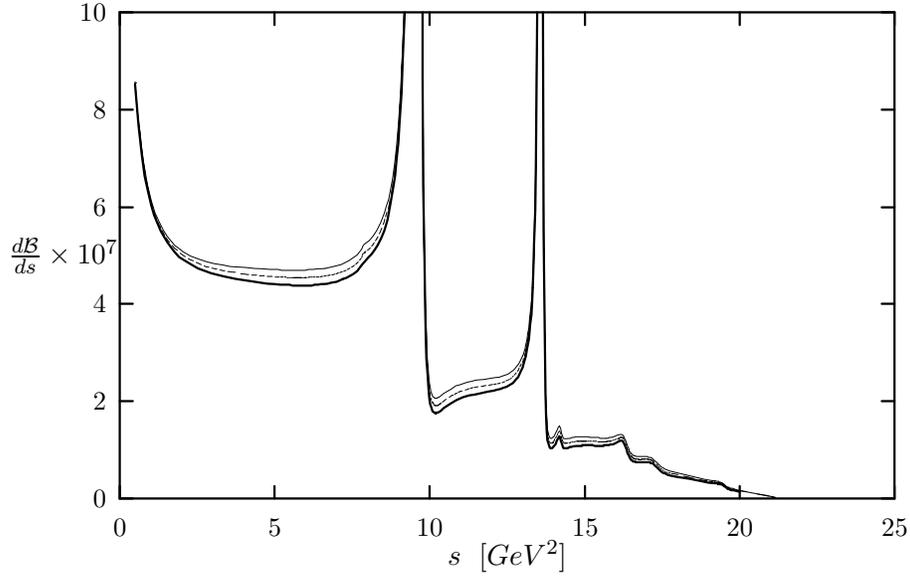
{Figure 6  ~~\small The differential branching ratio for the SM
(thick solid line), $\left|C_{LRLR}\right|^2 +
\left|C_{RLLR}\right|^2 + \left|C_{LRRL}\right|^2 +
\left|C_{RLRL}\right|^2 = 0.5 \left|C_{10}\right|^2 $ (thin
dashed line) and $\left|C_{10}\right|^2$ (thin solid line). } 
\bigskip 
\bigskip 
\bigskip 

\setlength{\unitlength}{0.240900pt}
%
 \\
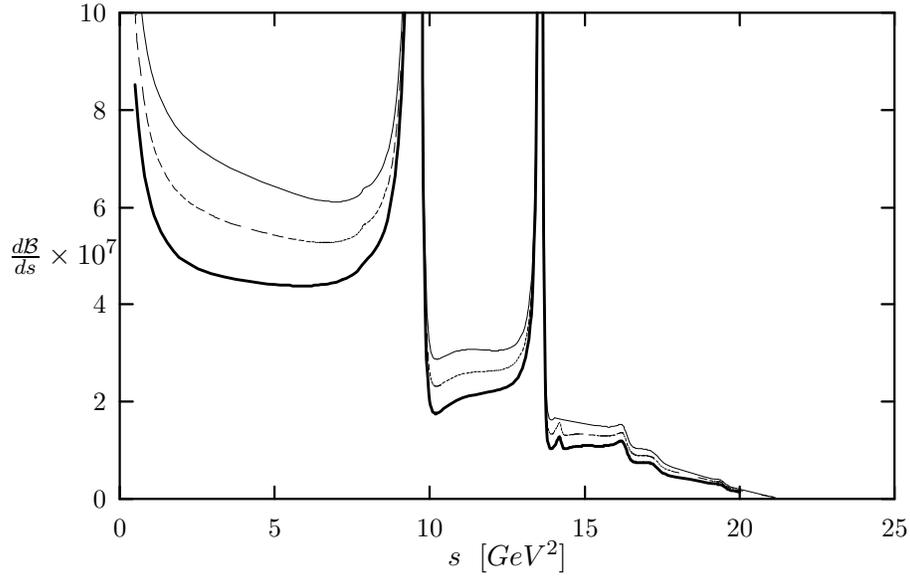
{Figure 7  ~~\small The differential branching ratio for the SM
(thick solid line), $16 \left|C_{T}\right|^2 + 64
\left|C_{TE}\right|^2 = 0.5 \left|C_{10}\right|^2 $ (thin
dashed line) and $\left|C_{10}\right|^2$ (thin solid line). } 

\bigskip
\bigskip
\bigskip

\bigskip 
\bigskip
\bigskip
\setlength{\unitlength}{0.240900pt}
%
 \\
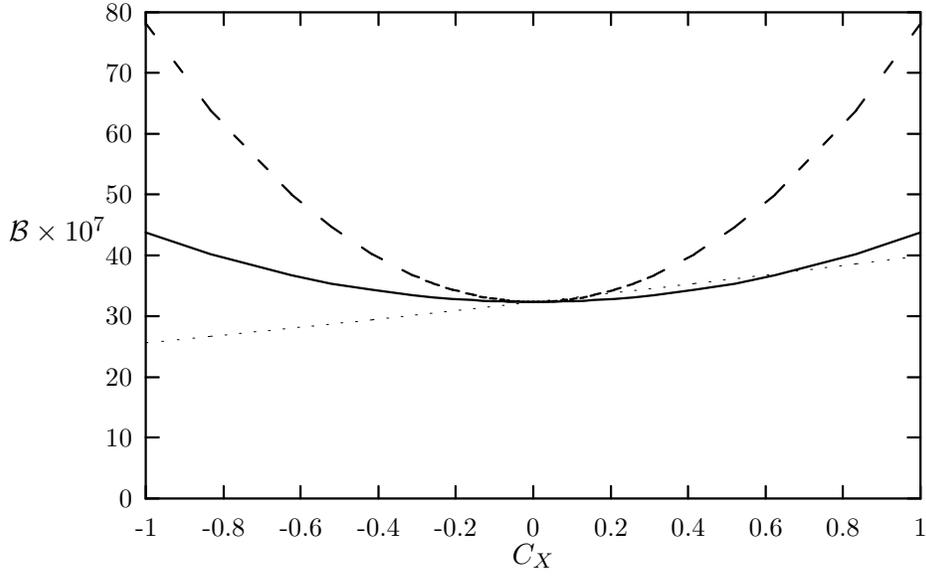
{Figure 8 \small The partially integrated branching ratio 
${\cal B}= \int_1^8\frac{ d {\cal B} }{ds}ds $ as a function of $C_{T}$ (thick
line), $C_{TE}$ (thick dashed line) and $C_{LL}$ (dotted 
dashed line). }

\bigskip
\bigskip
\bigskip
\bigskip
\bigskip

\setlength{\unitlength}{0.240900pt}
%
 \\
{Figure 9 \small The partially integrated branching ratio 
${\cal B}= \int_1^8\frac{ d {\cal B} }{ds}ds $ as a function of $C_{LL}$ 
(thick solid line), $C_{LR}$ (thick dashed line), $C_{RL}$ (thin
dashed line), $C_{RR}$( thin solid line ) and $C_{LRLR}$ (dotted line). }

\newpage

Now let us turn to the forward-backward (FB) asymmetry.
The normalized asymmetry $d{\bar {\cal A}}/ds$ is given by 
\bea
\frac{d\bar{{\cal A}}}{ds} = \frac{\frac{d{{\cal A}}}{ds}}
                           {\frac{d{{\cal B}}}{ds}}
     = \frac{ \int^{1}_0 dz \frac{d^2 {\cal B}}{dsdz}
             - \int^{0}_{-1} dz \frac{d^2 {\cal B}}{dsdz}}
            {\int^{1}_0 dz \frac{d^2 {\cal B}}{dsdz}
             + \int^{0}_{-1} dz \frac{d^2 {\cal B}}{dsdz}},
\eea
where $\frac{d{{\cal A}}}{ds}$ is the
FB  asymmetry and is expressed as
\bea
\frac{d{{\cal A}}}{ds} = \frac{1}{2 {m_b}^8 }{\cal B}_0  
             &[& A_3(s,1)~\{ \left|(C_9^{eff} - C_{10})\right|^2 
                         -  \left|(C_9^{eff} + C_{10})\right|^2 \} \nn \\
             &+& A_5(s,1)~\{ 2 Re(-2 C_7 ( C_9^{eff *} - C_{10}^* )) 
                          - 2 Re(-2 C_7 ( C_9^{eff *} + C_{10}^* )) 
                                                        \} \nn \\[5mm]
             &+& A_3(s,1)~\{ 2 Re(( C_9^{eff} - C_{10}){ C_{LL}^* })  
                          - 2 Re(( C_9^{eff} + C_{10}){ C_{LR}^* })
                                                            \} \nn \\
             &+& A_5(s,1)~\{ 2  Re( -2 C_7({C_{LL}^* - C_{LR}^* }))
                                                            \} \nn \\ 
             &+& A_7(s,1)~\{ 2 Re( -2 C_7({C_{RL}^* - C_{RR}^* })) 
                                                         \} \nn \\[5mm]
             &+& A_3(s,1)~\{ {  \left|C_{LL}\right|^2 
                                - \left|C_{LR}\right|^2 
                                -  \left|C_{RL}\right|^2 
                                + \left|C_{RR}\right|^2 } \} \nn \\
             &+& A_3(s,1)~\{ - 4 { Re[ C_{LRLR} (C_{T}^* - 2 C_{TE}^* )
                                  + C_{RLRL} (C_{T}^* + 2 C_{TE}^* )]}
                                                               \} ]. 
\eea
$A_n(s,1)$ are the functions which are obtained by integrating
  $A_n(s,z)$ with respect to $z$,
\bea
A_n(s,1) =  \int_{0}^1 A_n(s,z) dz - \int_{-1}^0 A_n(s,z) dz , 
\eea
\bea
A_3(s,1) &=& 2 u(s)^2 s ,\\
A_5(s,1) &=& - 2 ( m_b^2 + m_s^2 ) u(s)^2,  \\
A_7(s,1) &=& 4 m_b m_s u(s)^2,
\eea
where $n$ is 3, 5 and 7.
%
These kinematic functions are plotted
in Figure 10.  Among them, $A_7$ is smaller than the others
because it is suppressed by a factor of $m_s$.
In the same way as we have done for the differential rate, we plot the
normalized asymmetry by changing a coefficient of
vector type interactions within the range $-C_{10}< C_X < C_{10}$.  
The results are shown in Figures 11-14. 
These plots show that the asymmetry is the most sensitive to 
$C_{LL}$, which gives a positive (negative) contribution  for
$C_{LL}>0~ (<0)$. Figures 13 and 14 tell us that
$C_{RL}$ ($C_{RR}$) decreases (increases) the asymmetry.
To interpret the results, we collect the
terms coming from a vector type interaction 
in the FB asymmetry,    
\bea
A_3(s) \{   C_{LL}^2 + 2 ( C_9^{eff} - C_{10} ) C_{LL} \}  
                                           - 4 A_5(s) C_7 C_{LL}, \\
A_3(s) \{ - C_{LR}^2 - 2 ( C_9^{eff} + C_{10} ) C_{LR} \} 
                                           + 4 A_5(s) C_7 C_{LR}, \\
A_3(s) \{ - C_{RL}^2 \}, \\
A_3(s) \{   C_{RR}^2   \}, 
\eea
where we neglect the term proportional to $A_7$.
Because $(C_9^{eff} - C_{10})$ is much larger than the
other SM coefficients, the interference term with $C_{LL}$
is important. Therefore the asymmetry is sensitive to
$C_{LL}$. The response to  $C_{LR}$ is more involved 
and the interference term between  $C_7$ and $C_{LR}$
seems to be the most important. It gives a positive (negative)
contribution to the asymmetry for  $C_{LR}>0~ (<0)$.
We also note that $C_{RL}$ ($C_{RR}$)
gives a negative (positive) contribution to the asymmetry
and it is consistent
with the results shown in Figures 13 and 14.

We now study the integrated FB asymmetry (${\cal A}$).
We choose the same integration region as was chosen for the branching
ratio (${\cal B}$), and
define the integrated FB asymmetry ${\cal A}$
and the normalized FB asymmetry $\bar {\cal A}$,
\bea 
{\cal A}&=&\int_1^8 \frac{d {\cal A} }{d s}ds, \\ 
{\bar {\cal A} } &=& \int_1^8 \frac{d {\cal A} }{d s}ds/{\cal B},
\eea
where ${\cal B}= \int_1^8 \frac{d {\cal B} }{d s}ds$.
We outline how ${\cal A}$ and 
 $\bar {\cal A}$ depend on the
interactions. In Figure 15 (16), the dependence of the integrated asymmetry
${\cal A}$ (normalized asymmetry ${\bar {\cal A}}$ ) 
on $|C_{X}|$ is shown by changing the
coefficient $|C_{X}|$ within the small range, 
 $-C_{10}< |C_{X}| < C_{10}$. Within this range, the linear 
dependence on the coefficients is important.
At first glance, 
we see that only $C_{LL}$ can give rise to both large
positive and negative
contributions. $C_{LR}$ also gives rise to both positive and negative
contributions. 
$C_{RR}$ gives a positive contribution and 
$C_{RL}$ gives a negative contribution as already discussed 
for  $d{\cal A}/ds$.
We also note that  
the integrated FB asymmetry does not change 
at all as we change a coefficient of 
one of the  scalar or tensor interactions.  (See Figure 15.)
This is because the dependence 
on these types of interactions comes only from the interference
term. On the other hand, the magnitude of the
averaged  FB asymmetry is reduced slightly
 if one of these interactions act.
 The reason is that 
 the denominator of the averaged  FB asymmetry,
${\cal B}$, always increases as
scalar and tensor interactions are switched on. 
\bigskip

\setlength{\unitlength}{0.240900pt}
 \\
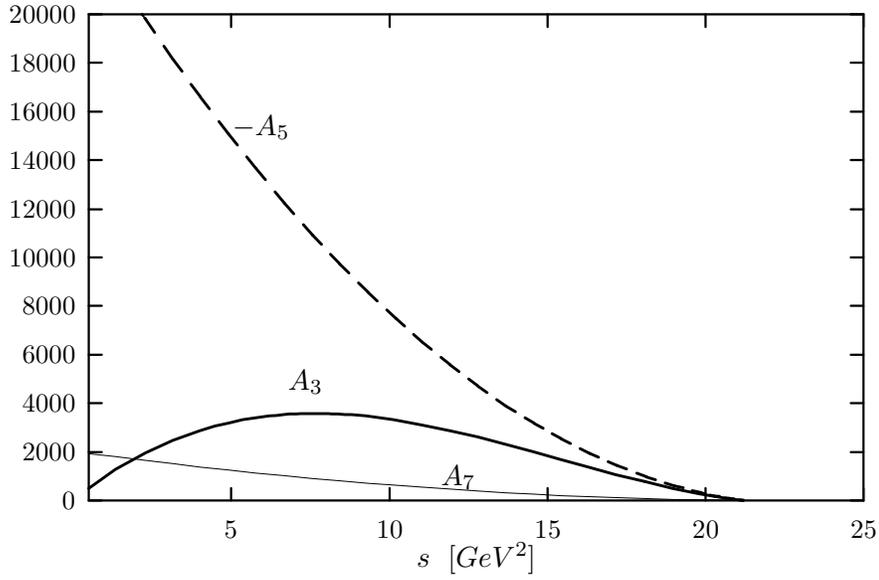
{Figure 10 ~\small  $A_3(s,1)$ (thin dotted line), 
$ -A_5(s,1)$ (thin solid line ) and 
$A_7(s,1)$ (thick solid line). }

\bigskip
\bigskip
\bigskip
\bigskip

\setlength{\unitlength}{0.240900pt}
%
 \\
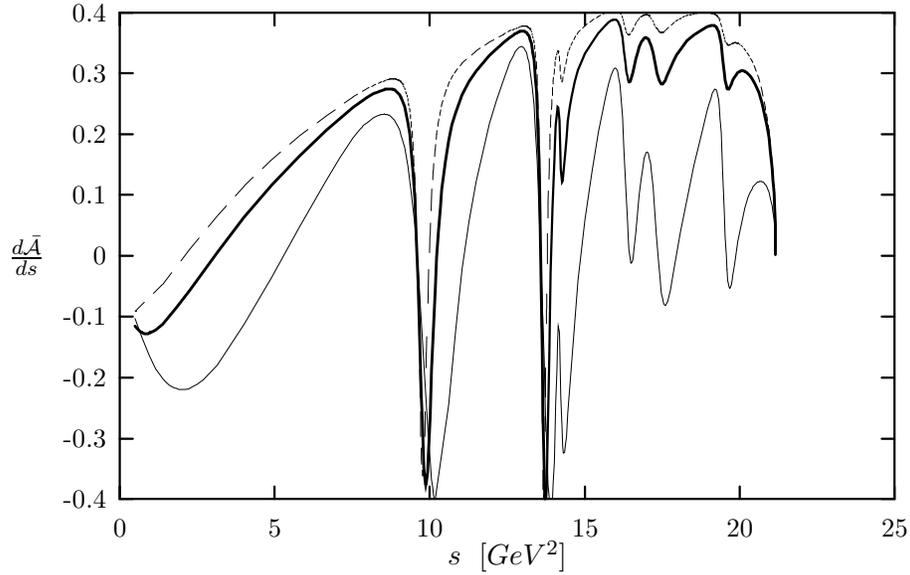
{Figuer 11 ~\small The normalized FB asymmetry for the SM (thick solid line),
$ C_{LL} $ is  
$ - \left|C_{10}\right|$ (thin line) 
and $ \left|C_{10}\right|$ (thin dashed line). The
coefficients of the other interactions in ${{\cal M}_{\rm NEW}}$
are set to zero.}

\bigskip
\bigskip

\setlength{\unitlength}{0.240900pt}
%
 \\
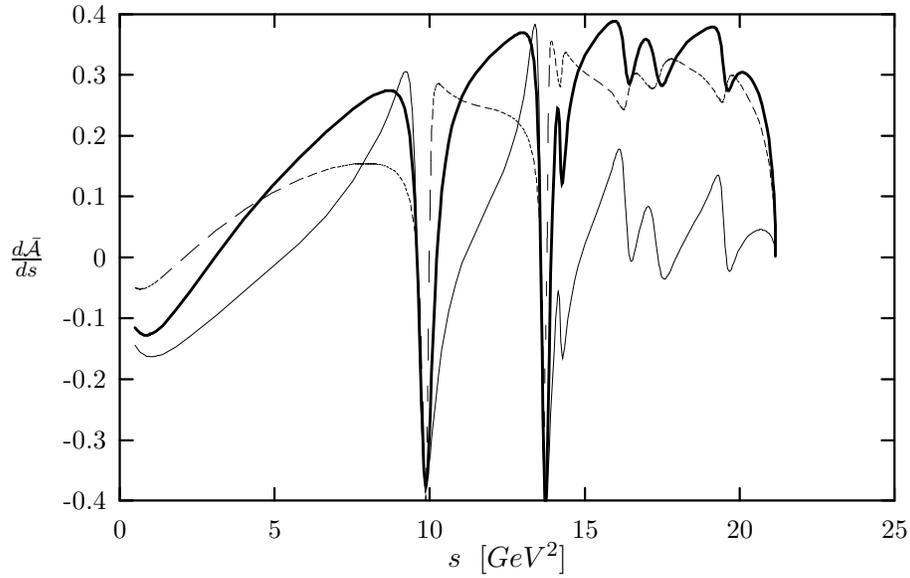
{Figure 12 \small The normalized FB asymmetry for the SM (thick solid line),
$ C_{LR} $ is  
$ - \left|C_{10}\right|$ (thin line) 
and $ \left|C_{10}\right|$ (thin dashed line). 
The coefficients of the other interactions in ${{\cal M}_{\rm NEW}}$
are set to zero.}

\bigskip
\bigskip
\bigskip
\bigskip

\setlength{\unitlength}{0.240900pt}
%
 \\
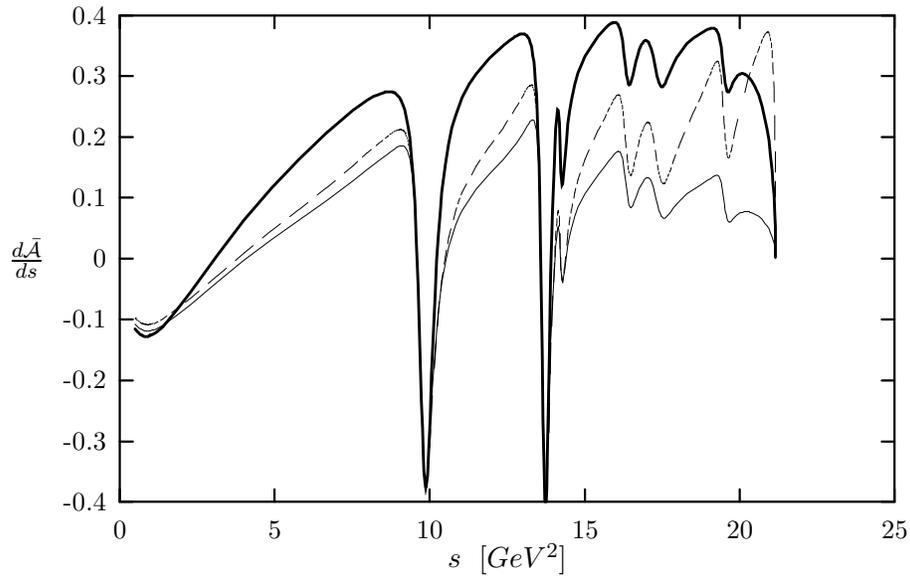
{Figure 13 \small The normalized FB asymmetry for the SM (thick solid line),
$ C_{RL} $ is  
$ - \left|C_{10}\right|$ (thin line) 
and $ \left|C_{10}\right|$ (thin dashed line).
The coefficients of the other interactions in ${{\cal M}_{\rm NEW}}$
are set to zero. }

\bigskip
\bigskip

\setlength{\unitlength}{0.240900pt}
%
 \\
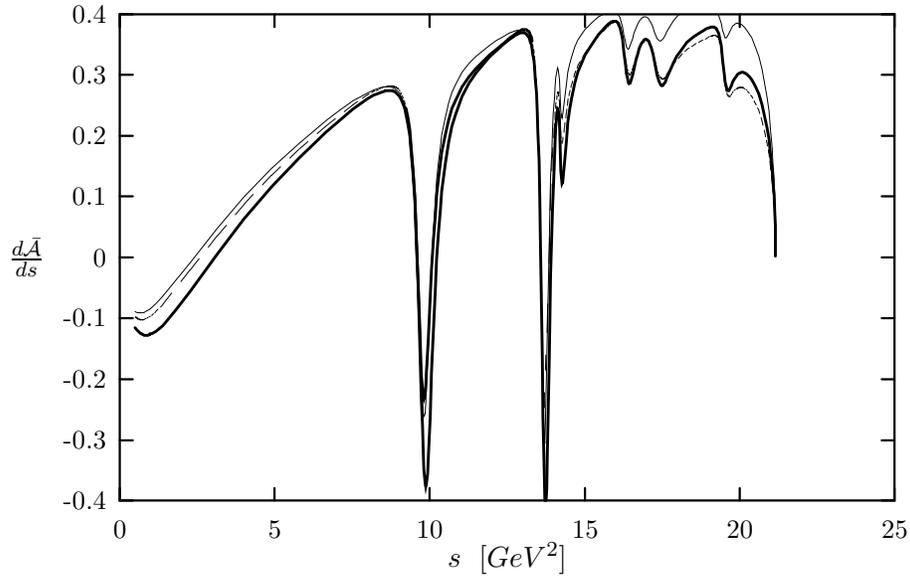
{Figure 14 \small The normalized FB asymmetry for the SM (thick solid line),
$ C_{RR} $ is  
$ - \left|C_{10}\right|$ (thin line) 
and $ - \left|C_{10}\right|$ (thin dashed line).
The coefficients of the other interactions in ${{\cal M}_{\rm NEW}}$
are set to zero.}
\bigskip
\bigskip
\bigskip
\bigskip

\bigskip

\setlength{\unitlength}{0.240900pt}
%
 \\
{Figure 15  ~\small The contribution to the integrated FB asymmetry ${\cal A}$
from $C_{LL}$ (thick
solid line), $C_{LR}$ (thick dashed line), $C_{RL}$ (thin
dashed line), $C_{RR}$ (thin solid line) and $C_{LRLR}$
(dotted line)}.
\bigskip

\setlength{\unitlength}{0.240900pt}
%
 \\
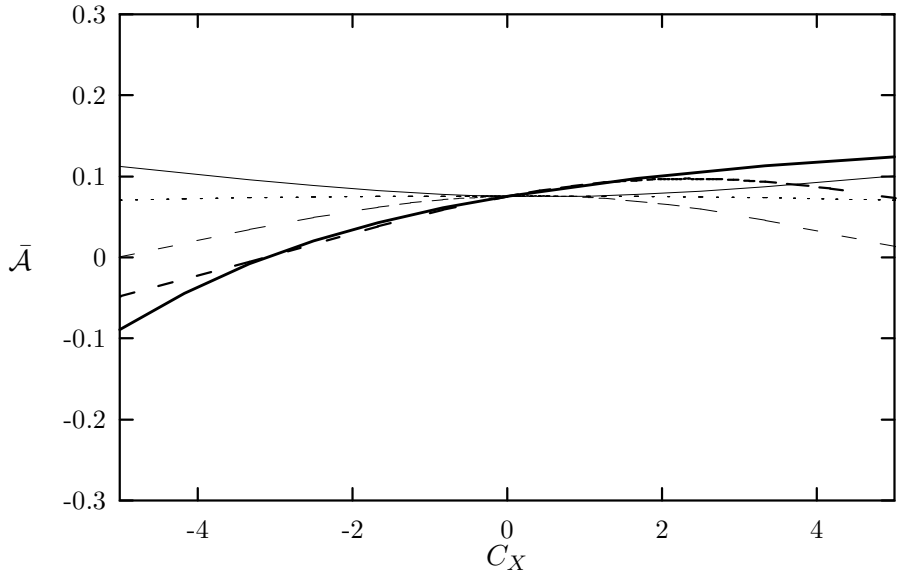
{Figure 16  ~\small The contribution to the integrated normalized FB asymmetry
$\bar{ {\cal A} }$ from $C_{LL}$ (thick
solid line), $C_{LR}$ (thick dashed line), $C_{RL}$ (thin
dashed line), $C_{RR}$ (thin solid line) and  $C_{L RLR}$
(dotted line)}.


\section{ Correlation between the branching ratio and the
forward-backward asymmetry}

In the previous Sections we have studied how the 
branching ratios and the asymmetries depend on each
interaction. In order to distinguish which interaction
contributes to the process, we plot the correlation
between the branching ratio and the asymmetry.
By changing a coefficient, we can draw a flow
in the two dimensional planes $({\cal B}, {\cal A})$ and $({\cal B}, 
{\bar {\cal A}})$.
The flows depend on  the type of interaction which acts.
Then we may pin down the type of interaction which
contributes to the processes 
once these quantities 
are measured. 
We allow the coefficients to vary in a wider range, {\it i.e.},
$-20<C_{X}<20$. The correlation plots  are shown for $({\cal B}, {\cal A})$
in Figure 17, and for $({\cal B}, {\bar {\cal A}})$ in Figure 18.
 ${\cal B}$, ${\cal A}$, 
and $\bar {\cal A}$
are also plotted as functions of the various coefficients in the 
range. (See Figures 19-21.)
 There are uncertainties in the SM
predictions which come from the renormalization point ($\mu$) dependence
of QCD correction. 
The predictions of the SM are shown for three values of
thr renormalization point.
The flows in the (${\cal B},{\cal A}$) plane
near the point of the SM prediction
 is summarized as follows.
\begin{itemize}
\item{
The figures  show that the flow for $C_{LL}$ is
distinctive from the others. In particular, if $C_{LL}$ 
is negative, the branching ratio decreases substantially while the other
interactions tend to increase the branching ratio.}
\item { As $|C_{LR}|$ and $|C_{RL}|$ increase, the asymmetry
decreases  except the
 range $0<C_{LR}<3$.  The branching ratio tends do
increases as  $|C_{LR}|$ and  $|C_{RL}|$ are larger.}
\item { As $|C_{RR}|$ increases, 
 the asymmetry and the branching ratio increase.}
\item{The tensor and scalar interactions do not affect the
asymmetry and increase the branching ratio. Therefore 
the flow becomes flat.}
\item{ The branching ratio and the asymmetry does not
strongly depend on the sign of  $C_{RL}$ and $C_{RR}$.
Therefore, the flow for the positive coefficent and that of
the negative coefficent are nearly 
degenerate. 
The flows for the positive $C_{LL}$ and $C_{LR}$
are distinct from  those for the negative coefficients.
They are not degenerate  because the branching ratio and/or the
asymmetry depend on their sign.}
\end{itemize}

The flow in the plane $({\cal B}, {\bar {\cal A}})$ is similar to that 
in  $({\cal B},  {\cal A})$ for small changes of the coefficients.
As coefficients are increased, nonlinear dependence begins
to be important. This causes the difference 
between ${\cal A}$ and ${\bar {\cal A}}$.
As an example, increasing the coefficient of
a  scalar or tensor operator, in the $({\cal B}, {\bar {\cal A}})$ plane, 
the asymptote of the flow is ${\bar {\cal A}}=0$, 
while in the $({\cal B}, {\cal A})$
plane, it is flat. 
\bigskip

\setlength{\unitlength}{0.240900pt}
\begin{picture}(1500,900)(0,0)
\tenrm
\thicklines \path(220,113)(240,113)
\thicklines \path(1436,113)(1416,113)
\put(198,113){\makebox(0,0)[r]{-15}}
\thicklines \path(220,240)(240,240)
\thicklines \path(1436,240)(1416,240)
\put(198,240){\makebox(0,0)[r]{-10}}
\thicklines \path(220,368)(240,368)
\thicklines \path(1436,368)(1416,368)
\put(198,368){\makebox(0,0)[r]{-5}}
\thicklines \path(220,495)(240,495)
\thicklines \path(1436,495)(1416,495)
\put(198,495){\makebox(0,0)[r]{0}}
\thicklines \path(220,622)(240,622)
\thicklines \path(1436,622)(1416,622)
\put(198,622){\makebox(0,0)[r]{5}}
\thicklines \path(220,750)(240,750)
\thicklines \path(1436,750)(1416,750)
\put(198,750){\makebox(0,0)[r]{10}}
\thicklines \path(220,877)(240,877)
\thicklines \path(1436,877)(1416,877)
\put(198,877){\makebox(0,0)[r]{15}}
\thicklines \path(220,113)(220,133)
\thicklines \path(220,877)(220,857)
\put(220,68){\makebox(0,0){0}}
\thicklines \path(463,113)(463,133)
\thicklines \path(463,877)(463,857)
\put(463,68){\makebox(0,0){20}}
\thicklines \path(706,113)(706,133)
\thicklines \path(706,877)(706,857)
\put(706,68){\makebox(0,0){40}}
\thicklines \path(950,113)(950,133)
\thicklines \path(950,877)(950,857)
\put(950,68){\makebox(0,0){60}}
\thicklines \path(1193,113)(1193,133)
\thicklines \path(1193,877)(1193,857)
\put(1193,68){\makebox(0,0){80}}
\thicklines \path(1436,113)(1436,133)
\thicklines \path(1436,877)(1436,857)
\put(1436,68){\makebox(0,0){100}}
\thicklines \path(220,113)(1436,113)(1436,877)(220,877)(220,113)
\put(25,455){\makebox(0,0)[l]{\shortstack{$ {\cal A} \times 10^7 $}}}
\put(828,23){\makebox(0,0){$ {\cal B} \times 10^7 $}}
\put(450,300){\makebox(0,0)[lb]{$ \bigtriangleup ~\mu = 10({\rm GeV}) $}}
\put(450,240){\makebox(0,0)[lb]{$ \bigcirc ~\mu = 5 ({\rm GeV}) $}}
\put(450,180){\makebox(0,0)[lb]{$ \bigtriangledown ~\mu = 2.5({\rm GeV})$}}
\Thicklines \path(967,877)(817,804)(649,721)(511,649)(403,590)(325,543)(297,525)(277,509)(264,496)(260,491)(258,486)(258,483)(260,480)(264,477)(270,476)(277,475)(286,476)(311,478)(343,484)(382,492)(483,519)(613,557)(774,608)(964,671)(1184,746)(1435,834)(1436,834)
\thicklines \dashline[-10]{18}(1436,119)(1253,206)(1072,295)(921,372)(799,437)(707,489)(673,511)(645,529)(626,545)(613,557)(610,562)(608,567)(609,570)(611,573)(613,574)(615,575)(621,576)(629,577)(639,577)(664,574)(696,568)(783,547)(901,514)(1047,469)(1224,411)(1431,342)(1436,340)
\thinlines \dashline[-10]{18}(1436,225)(1392,243)(1192,325)(1021,394)(879,451)(768,496)(724,514)(687,529)(657,540)(635,549)(627,552)(621,555)(618,556)(616,556)(615,557)(614,557)(614,557)(613,557)(613,557)(613,557)(613,557)(613,557)(613,557)(613,557)(613,557)(613,557)(613,557)(613,557)(613,557)(613,557)(614,557)(615,556)(617,555)(621,553)(628,550)(637,547)(659,537)(689,525)(727,509)(825,468)(952,415)(1109,350)(1296,273)(1436,216)
\thinlines \path(1357,877)(1344,872)(1150,790)(986,721)(852,664)(748,619)(707,601)(673,586)(647,574)(628,566)(622,562)(617,560)(615,559)(614,558)(614,558)(614,558)(613,557)(613,557)(613,557)(613,557)(613,557)(614,557)(615,557)(617,558)(619,558)(622,559)(628,561)(637,564)(647,568)(673,577)(707,590)(748,606)(852,646)(986,699)(1151,764)(1345,841)(1434,877)
\thinlines \dashline[-10]{5}(1036,557)(1036,557)(968,557)(907,557)(851,557)(801,557)(757,557)(719,557)(687,557)(660,557)(640,557)(632,557)(625,557)(620,557)(618,557)(616,557)(615,557)(613,557)(616,557)(618,557)(640,557)(687,557)(1036,557)(1193,557)(1436,557)
\thicklines \path(676,600)(676,600)(613,557)(561,508)
\put(676,600){\raisebox{-1.2pt}{\makebox(0,0){$\small \bigtriangleup $}}}
\put(613,557){\raisebox{-1.2pt}{\makebox(0,0){$\large \circ $}}}
\put(561,508){\raisebox{-1.2pt}{\makebox(0,0){$\small \bigtriangledown $}}}
\end{picture} \\
{Figure 17 ~\small  The flows in $({\cal B},  {\cal A})$ plane.
In each flow,  $C_{LL}$ (thick
solid line ), $C_{LR}$ (thick dashed line), $C_{RL}$ (thin
dashed line), $C_{RR}$ (thin solid line),
 and the tensor and scalar interactions (dotted line)
are changed respectively. 
The predictions of the SM
 are denoted by $\small \bigtriangleup $, $\Large \circ $, and $\small
\bigtriangledown $  corresponding to
 three values of the renormalization point ($\mu$).}

\bigskip
\bigskip

\setlength{\unitlength}{0.240900pt}
\begin{picture}(1500,900)(0,0)
\tenrm
\thicklines \path(220,113)(240,113)
\thicklines \path(1436,113)(1416,113)
\put(198,113){\makebox(0,0)[r]{-0.2}}
\thicklines \path(220,208)(240,208)
\thicklines \path(1436,208)(1416,208)
\put(198,208){\makebox(0,0)[r]{-0.15}}
\thicklines \path(220,304)(240,304)
\thicklines \path(1436,304)(1416,304)
\put(198,304){\makebox(0,0)[r]{-0.1}}
\thicklines \path(220,400)(240,400)
\thicklines \path(1436,400)(1416,400)
\put(198,400){\makebox(0,0)[r]{-0.05}}
\thicklines \path(220,495)(240,495)
\thicklines \path(1436,495)(1416,495)
\put(198,495){\makebox(0,0)[r]{0}}
\thicklines \path(220,591)(240,591)
\thicklines \path(1436,591)(1416,591)
\put(198,591){\makebox(0,0)[r]{0.05}}
\thicklines \path(220,686)(240,686)
\thicklines \path(1436,686)(1416,686)
\put(198,686){\makebox(0,0)[r]{0.1}}
\thicklines \path(220,782)(240,782)
\thicklines \path(1436,782)(1416,782)
\put(198,782){\makebox(0,0)[r]{0.15}}
\thicklines \path(220,877)(240,877)
\thicklines \path(1436,877)(1416,877)
\put(198,877){\makebox(0,0)[r]{0.2}}
\thicklines \path(220,113)(220,133)
\thicklines \path(220,877)(220,857)
\put(220,68){\makebox(0,0){0}}
\thicklines \path(463,113)(463,133)
\thicklines \path(463,877)(463,857)
\put(463,68){\makebox(0,0){20}}
\thicklines \path(706,113)(706,133)
\thicklines \path(706,877)(706,857)
\put(706,68){\makebox(0,0){40}}
\thicklines \path(950,113)(950,133)
\thicklines \path(950,877)(950,857)
\put(950,68){\makebox(0,0){60}}
\thicklines \path(1193,113)(1193,133)
\thicklines \path(1193,877)(1193,857)
\put(1193,68){\makebox(0,0){80}}
\thicklines \path(1436,113)(1436,133)
\thicklines \path(1436,877)(1436,857)
\put(1436,68){\makebox(0,0){100}}
\thicklines \path(220,113)(1436,113)(1436,877)(220,877)(220,113)
\put(45,495){\makebox(0,0)[l]{\shortstack{$ \bar{\cal A} $}}}
\put(828,23){\makebox(0,0){$ {\cal B} \times 10^7  $}}
\put(550,300){\makebox(0,0)[lb]{$ \bigtriangleup ~\mu = 10({\rm GeV})  $}}
\put(550,240){\makebox(0,0)[lb]{$ \bigcirc ~\mu = 5({\rm GeV})$}}
\put(550,180){\makebox(0,0)[lb]{$ \bigtriangledown ~\mu = 2.5({\rm GeV}) $}}
\Thicklines \path(307,877)(303,866)(297,844)(291,818)
(286,788)(281,754)(277,716)(273,672)(269,624)(266,572)(264,516)
(262,457)(260,399)(259,341)(259,314)(258,301)
(258,288)(258,275)(258,263)(258,251)(258,240)
(258,229)(258,219)(258,209)(258,200)(258,191)(259,183)
(259,169)(259,162)(260,157)
(260,152)(260,147)(261,143)(261,140)(261,137)(262,135)
(262,133)(263,132)(263,132)
(264,132)(265,132)(265,133)(267,136)(268,141)(270,147)
(286,228)(298,277)(311,325)
(326,370)(343,410)(382,480)
\Thicklines \path(382,480)(404,508)(428,534)(483,577)(513,595)
(544,612)(613,639)(690,662)(774,681)(865,697)(964,711)(1071,723)
(1184,733)(1306,742)(1435,749)(1436,750)
\thicklines \dashline[-10]{18}(1436,213)(1253,240)(1072,281)(921,335)(799,403)(707,484)(645,569)(626,607)(613,639)(610,653)(608,663)(609,672)(610,675)(611,677)(613,679)(615,680)(621,680)(629,678)(639,673)(664,657)(696,635)(783,579)(901,521)(1047,466)(1224,419)(1431,380)(1436,379)
\thinlines \dashline[-10]{18}(1436,294)(1392,299)(1192,335)(1021,380)(879,434)(768,496)(687,561)(657,590)(635,614)(627,623)(621,631)(618,634)(616,636)(615,637)(614,638)(614,639)(613,639)(613,640)(613,640)(613,640)(613,640)(613,640)(613,640)(613,640)(613,640)(613,640)(613,640)(613,639)(613,639)(614,638)(615,636)(617,634)(621,628)(628,619)(637,608)(659,583)(727,520)(825,455)(952,396)(1109,347)(1296,307)(1436,287)
\thinlines \path(1436,806)(1344,800)(1150,784)(986,764)(852,738)(748,709)(673,678)(647,664)(628,653)(622,648)(617,644)(615,643)(614,641)(614,640)(613,640)(613,639)(613,639)(613,639)(614,639)(614,639)(614,639)(615,639)(615,639)(617,639)(619,640)(622,641)(628,643)(647,651)(673,661)(748,686)(852,713)(986,738)(1151,759)(1345,776)(1436,782)
\thinlines \dashline[-10]{5}(1436,542)(1213,552)(997,568)(829,588)(709,611)(667,622)(637,631)(627,635)(619,637)(617,638)(615,639)(614,639)(613,639)(614,639)(615,639)(617,638)(619,637)(627,635)(637,631)(667,622)(709,611)(829,588)(997,568)(1213,552)(1436,542)
\thicklines \path(676,706)(676,706)(613,639)(561,530)
\put(676,706){\raisebox{-1.2pt}{\makebox(0,0){$\small \bigtriangleup $}}}
\put(613,639){\raisebox{-1.2pt}{\makebox(0,0){$\Large \circ $}}}
\put(561,530){\raisebox{-1.2pt}{\makebox(0,0){$\small \bigtriangledown $}}}
\end{picture} \\
{Figure 18 ~\small  The flows in $({\cal B}, {\bar {\cal A}})$ plane.
In each flow,
$C_{LL}$ (thick
solid line), $C_{LR}$ (thick dashed line), $C_{RL}$ (thin
dashed line), $C_{RR}$ (thin solid line), and the tensor
 and scalar interactions (dotted line) are changed respectively.
The predictions of the SM are denoted as
$\small \bigtriangleup $, $\Large \circ $, and $\small
\bigtriangledown $  corresponding to
 three values of the renormalization point ($\mu$).
}

\bigskip
\bigskip
\bigskip
\bigskip

\setlength{\unitlength}{0.240900pt}
\begin{picture}(1500,900)(0,0)
\tenrm
\thicklines \path(220,113)(240,113)
\thicklines \path(1436,113)(1416,113)
\put(198,113){\makebox(0,0)[r]{0}}
\thicklines \path(220,189)(240,189)
\thicklines \path(1436,189)(1416,189)
\put(198,189){\makebox(0,0)[r]{20}}
\thicklines \path(220,266)(240,266)
\thicklines \path(1436,266)(1416,266)
\put(198,266){\makebox(0,0)[r]{40}}
\thicklines \path(220,342)(240,342)
\thicklines \path(1436,342)(1416,342)
\put(198,342){\makebox(0,0)[r]{60}}
\thicklines \path(220,419)(240,419)
\thicklines \path(1436,419)(1416,419)
\put(198,419){\makebox(0,0)[r]{80}}
\thicklines \path(220,495)(240,495)
\thicklines \path(1436,495)(1416,495)
\put(198,495){\makebox(0,0)[r]{100}}
\thicklines \path(220,571)(240,571)
\thicklines \path(1436,571)(1416,571)
\put(198,571){\makebox(0,0)[r]{120}}
\thicklines \path(220,648)(240,648)
\thicklines \path(1436,648)(1416,648)
\put(198,648){\makebox(0,0)[r]{140}}
\thicklines \path(220,724)(240,724)
\thicklines \path(1436,724)(1416,724)
\put(198,724){\makebox(0,0)[r]{160}}
\thicklines \path(220,801)(240,801)
\thicklines \path(1436,801)(1416,801)
\put(198,801){\makebox(0,0)[r]{180}}
\thicklines \path(220,877)(240,877)
\thicklines \path(1436,877)(1416,877)
\put(198,877){\makebox(0,0)[r]{200}}
\thicklines \path(220,113)(220,133)
\thicklines \path(220,877)(220,857)
\put(220,68){\makebox(0,0){-20}}
\thicklines \path(372,113)(372,133)
\thicklines \path(372,877)(372,857)
\put(372,68){\makebox(0,0){-15}}
\thicklines \path(524,113)(524,133)
\thicklines \path(524,877)(524,857)
\put(524,68){\makebox(0,0){-10}}
\thicklines \path(676,113)(676,133)
\thicklines \path(676,877)(676,857)
\put(676,68){\makebox(0,0){-5}}
\thicklines \path(828,113)(828,133)
\thicklines \path(828,877)(828,857)
\put(828,68){\makebox(0,0){0}}
\thicklines \path(980,113)(980,133)
\thicklines \path(980,877)(980,857)
\put(980,68){\makebox(0,0){5}}
\thicklines \path(1132,113)(1132,133)
\thicklines \path(1132,877)(1132,857)
\put(1132,68){\makebox(0,0){10}}
\thicklines \path(1284,113)(1284,133)
\thicklines \path(1284,877)(1284,857)
\put(1284,68){\makebox(0,0){15}}
\thicklines \path(1436,113)(1436,133)
\thicklines \path(1436,877)(1436,857)
\put(1436,68){\makebox(0,0){20}}
\thicklines \path(220,113)(1436,113)(1436,877)(220,877)(220,113)
\put(5,535){\makebox(0,0)[l]{\shortstack{$ {\cal B} \times 10^7 $}}}
\put(828,23){\makebox(0,0){$ C_X $}}
\Thicklines \path(220,363)(220,363)(271,300)(321,248)(372,205)(423,171)(448,157)(473,146)(499,137)(511,134)(524,131)(537,128)(549,127)(556,126)(562,126)(565,125)(568,125)(571,125)(573,125)(575,125)(576,125)(578,125)(579,125)(581,125)(583,125)(584,125)(586,125)(587,125)(590,125)(594,125)(597,125)(600,126)(606,126)(613,127)(625,129)(638,131)(651,134)(676,142)(701,152)(727,164)(777,196)(828,237)(879,287)(929,347)(980,416)(1031,495)(1081,582)(1132,680)(1183,786)(1222,877)
\thicklines \dashline[-10]{25}(264,877)(271,863)(321,759)(372,665)(423,580)(473,504)(524,438)(575,381)(625,333)(676,295)(701,279)(727,266)(752,255)(777,247)(790,243)(803,240)(815,238)(822,237)(828,237)(834,236)(837,236)(841,236)(844,235)(847,235)(849,235)(850,235)(852,235)(853,235)(855,235)(856,235)(858,235)(860,235)(861,235)(863,235)(866,235)(868,235)(869,235)(872,235)(879,236)(885,236)(891,237)(904,239)(917,241)(929,245)(955,252)(980,263)(1031,290)(1081,327)(1132,373)(1183,428)(1233,493)(1233,493)(1284,568)(1335,651)(1385,744)(1436,847)
\thinlines \dashline[-10]{25}(248,877)(271,827)(321,727)(372,636)(423,554)(473,481)(524,418)(575,365)(625,320)(676,285)(701,271)(727,260)(752,250)(777,243)(790,241)(803,239)(809,238)(815,237)(818,237)(822,237)(825,237)(828,237)(830,237)(831,236)(833,236)(834,236)(836,236)(837,236)(839,236)(841,236)(842,236)(844,236)(847,236)(849,236)(850,237)(853,237)(860,237)(866,238)(879,239)(891,241)(904,244)(929,251)(955,260)(980,272)(1031,303)(1081,343)(1132,392)(1183,451)(1233,519)(1284,597)(1335,684)
\thinlines \dashline[-10]{25}(1335,684)(1385,780)(1432,877)
\thinlines \path(236,877)(271,804)(321,705)(372,616)(423,536)(473,466)(524,405)(575,354)(625,312)(676,279)(701,266)(727,255)(752,247)(765,244)(777,241)(790,239)(796,238)(803,238)(809,237)(812,237)(815,237)(818,237)(820,237)(822,237)(823,237)(825,237)(826,237)(828,237)(830,237)(831,237)(833,237)(834,237)(837,237)(841,237)(844,237)(847,237)(853,238)(860,238)(866,239)(879,241)(891,244)(904,247)(929,255)(955,266)(980,279)(1031,312)(1081,354)(1132,405)(1183,466)(1233,537)(1284,616)
\thinlines \path(1284,616)(1335,705)(1385,804)(1420,877)
\thinlines \dashline[-10]{5}(220,369)(220,369)(271,348)(321,329)(372,311)(423,296)(473,282)(524,270)(575,260)(625,251)(676,245)(701,242)(727,240)(752,239)(765,238)(777,237)(790,237)(796,237)(803,237)(809,237)(812,237)(815,237)(818,237)(820,237)(822,237)(823,237)(825,237)(826,237)(828,237)(830,237)(831,237)(833,237)(834,237)(837,237)(841,237)(844,237)(847,237)(853,237)(860,237)(866,237)(879,237)(891,238)(904,239)(929,240)(955,242)(980,245)(1031,251)(1081,260)(1132,270)(1183,282)(1233,296)(1233,296)(1284,311)(1335,329)(1385,348)(1436,369)
\thinlines \dashline[-10]{5}(220,369)(220,369)(271,348)(321,329)(372,311)(423,296)(473,282)(524,270)(575,260)(625,251)(676,245)(701,242)(727,240)(752,239)(765,238)(777,237)(790,237)(796,237)(803,237)(809,237)(812,237)(815,237)(818,237)(820,237)(822,237)(823,237)(825,237)(826,237)(828,237)(830,237)(831,237)(833,237)(834,237)(837,237)(841,237)(844,237)(847,237)(853,237)(860,237)(866,237)(879,237)(891,238)(904,239)(929,240)(955,242)(980,245)(1031,251)(1081,260)(1132,270)(1183,282)(1233,296)(1233,296)(1284,311)(1335,329)(1385,348)(1436,369)
\thinlines \dashline[-10]{5}(220,369)(220,369)(271,348)(321,329)(372,311)(423,296)(473,282)(524,270)(575,260)(625,251)(676,245)(701,242)(727,240)(752,239)(765,238)(777,237)(790,237)(796,237)(803,237)(809,237)(812,237)(815,237)(818,237)(820,237)(822,237)(823,237)(825,237)(826,237)(828,237)(830,237)(831,237)(833,237)(834,237)(837,237)(841,237)(844,237)(847,237)(853,237)(860,237)(866,237)(879,237)(891,238)(904,239)(929,240)(955,242)(980,245)(1031,251)(1081,260)(1132,270)(1183,282)(1233,296)
\thinlines \dashline[-10]{5}(1233,296)(1284,311)(1335,329)(1385,348)(1436,369)
\thinlines \dashline[-10]{5}(220,369)(220,369)(271,348)(321,329)(372,311)(423,296)(473,282)(524,270)(575,260)(625,251)(676,245)(701,242)(727,240)(752,239)(765,238)(777,237)(790,237)(796,237)(803,237)(809,237)(812,237)(815,237)(818,237)(820,237)(822,237)(823,237)(825,237)(826,237)(828,237)(830,237)(831,237)(833,237)(834,237)(837,237)(841,237)(844,237)(847,237)(853,237)(860,237)(866,237)(879,237)(891,238)(904,239)(929,240)(955,242)(980,245)(1031,251)(1081,260)(1132,270)(1183,282)(1233,296)
\thinlines \dashline[-10]{5}(1233,296)(1284,311)(1335,329)(1385,348)(1436,369)
\end{picture}\\
{Figure 19~ \small The partially integrated branching ratio 
${\cal B}= \int_1^8\frac{ d 
{\cal B} }{ds}ds $ as a function of $C_{LL}$ (thick
solid line), $C_{LR}$ (thick dashed line), $C_{RL}$ (thin
dashed line), $C_{RR}$ (thin solid line) and $C_{LRLR}$ (dotted line). }

\bigskip
\bigskip

\setlength{\unitlength}{0.240900pt}
\begin{picture}(1500,900)(0,0)
\tenrm
\thicklines \path(220,113)(240,113)
\thicklines \path(1436,113)(1416,113)
\put(198,113){\makebox(0,0)[r]{-15}}
\thicklines \path(220,240)(240,240)
\thicklines \path(1436,240)(1416,240)
\put(198,240){\makebox(0,0)[r]{-10}}
\thicklines \path(220,368)(240,368)
\thicklines \path(1436,368)(1416,368)
\put(198,368){\makebox(0,0)[r]{-5}}
\thicklines \path(220,495)(240,495)
\thicklines \path(1436,495)(1416,495)
\put(198,495){\makebox(0,0)[r]{0}}
\thicklines \path(220,622)(240,622)
\thicklines \path(1436,622)(1416,622)
\put(198,622){\makebox(0,0)[r]{5}}
\thicklines \path(220,750)(240,750)
\thicklines \path(1436,750)(1416,750)
\put(198,750){\makebox(0,0)[r]{10}}
\thicklines \path(220,877)(240,877)
\thicklines \path(1436,877)(1416,877)
\put(198,877){\makebox(0,0)[r]{15}}
\thicklines \path(220,113)(220,133)
\thicklines \path(220,877)(220,857)
\put(220,68){\makebox(0,0){-20}}
\thicklines \path(372,113)(372,133)
\thicklines \path(372,877)(372,857)
\put(372,68){\makebox(0,0){-15}}
\thicklines \path(524,113)(524,133)
\thicklines \path(524,877)(524,857)
\put(524,68){\makebox(0,0){-10}}
\thicklines \path(676,113)(676,133)
\thicklines \path(676,877)(676,857)
\put(676,68){\makebox(0,0){-5}}
\thicklines \path(828,113)(828,133)
\thicklines \path(828,877)(828,857)
\put(828,68){\makebox(0,0){0}}
\thicklines \path(980,113)(980,133)
\thicklines \path(980,877)(980,857)
\put(980,68){\makebox(0,0){5}}
\thicklines \path(1132,113)(1132,133)
\thicklines \path(1132,877)(1132,857)
\put(1132,68){\makebox(0,0){10}}
\thicklines \path(1284,113)(1284,133)
\thicklines \path(1284,877)(1284,857)
\put(1284,68){\makebox(0,0){15}}
\thicklines \path(1436,113)(1436,133)
\thicklines \path(1436,877)(1436,857)
\put(1436,68){\makebox(0,0){20}}
\thicklines \path(220,113)(1436,113)(1436,877)(220,877)(220,113)
\put(25,525){\makebox(0,0)[l]{\shortstack{$ {\cal A} \times 10^7  $}}}
\put(828,23){\makebox(0,0){$ C_X $}}
\Thicklines \path(232,877)(271,804)(321,721)(372,649)(423,590)(473,543)(499,525)(524,509)(549,496)(562,491)(575,486)(587,483)(600,480)(613,477)(619,477)(622,476)(625,476)(628,476)(632,476)(633,476)(635,476)(636,475)(638,475)(640,475)(641,475)(643,475)(644,475)(646,475)(647,475)(651,476)(652,476)(654,476)(657,476)(663,476)(670,477)(676,478)(689,480)(701,484)(727,492)(752,504)(777,519)(828,557)(879,608)(929,671)(980,746)(1031,834)(1053,877)
\thicklines \dashline[-10]{25}(477,113)(524,206)(575,295)(625,372)(676,437)(727,489)(752,511)(777,529)(803,545)(828,557)(841,562)(853,567)(866,570)(879,573)(885,574)(891,575)(898,576)(901,576)(904,576)(907,577)(909,577)(910,577)(912,577)(913,577)(915,577)(917,577)(918,577)(920,577)(921,577)(923,577)(926,577)(928,577)(929,577)(932,576)(936,576)(942,576)(948,575)(955,574)(967,571)(980,568)(1005,559)(1031,547)(1081,514)(1132,469)(1183,411)(1233,342)(1284,260)(1335,166)(1360,113)
\thinlines \dashline[-10]{25}(405,113)(423,149)(473,243)(524,325)(575,394)(625,451)(676,496)(701,514)(727,529)(752,540)(777,549)(790,552)(803,555)(809,556)(815,556)(818,557)(822,557)(825,557)(826,557)(828,557)(830,557)(831,557)(833,557)(834,557)(836,557)(837,557)(839,557)(841,557)(842,557)(844,557)(847,557)(850,557)(853,557)(860,556)(866,555)(879,553)(891,550)(904,547)(929,537)(955,525)(980,509)(1031,468)(1081,415)(1132,350)(1183,273)(1233,184)(1269,113)
\thinlines \path(470,877)(473,872)(524,790)(575,721)(625,664)(676,619)(701,601)(727,586)(752,574)(777,566)(790,562)(803,560)(809,559)(815,558)(818,558)(822,558)(825,557)(826,557)(828,557)(830,557)(831,557)(833,557)(834,557)(836,557)(837,557)(839,557)(841,557)(842,557)(844,557)(847,557)(850,558)(853,558)(860,558)(866,559)(879,561)(891,564)(904,568)(929,577)(955,590)(980,606)(1031,646)(1081,699)(1132,764)(1183,841)(1203,877)
\thinlines \dashline[-10]{5}(220,557)(220,557)(271,557)(321,557)(372,557)(423,557)(473,557)(524,557)(575,557)(625,557)(676,557)(727,557)(777,557)(828,557)(879,557)(929,557)(980,557)(1031,557)(1081,557)(1132,557)(1183,557)(1233,557)(1284,557)(1335,557)(1385,557)(1436,557)
\thinlines \dashline[-10]{5}(220,557)(220,557)(271,557)(321,557)(372,557)(423,557)(473,557)(524,557)(575,557)(625,557)(676,557)(727,557)(777,557)(828,557)(879,557)(929,557)(980,557)(1031,557)(1081,557)(1132,557)(1183,557)(1233,557)(1284,557)(1335,557)(1385,557)(1436,557)
\thinlines \dashline[-10]{5}(220,557)(220,557)(271,557)(321,557)(372,557)(423,557)(473,557)(524,557)(575,557)(625,557)(676,557)(727,557)(777,557)(828,557)(879,557)(929,557)(980,557)(1031,557)(1081,557)(1132,557)(1183,557)(1233,557)(1284,557)(1335,557)(1385,557)(1436,557)
\end{picture} \\
{Figure 20  ~\small The contribution to the integrated FB asymmetry ${\cal A}$
from $C_{LL}$ (thick
solid line), $C_{LR}$ (thick dashed line), $C_{RL}$ (thin
dashed line), $C_{RR}$ (thin solid line) and the tensor
and scalar interactions  (dotted line). }
\bigskip
\bigskip
\bigskip

\setlength{\unitlength}{0.240900pt}
\begin{picture}(1500,900)(0,0)
\tenrm
\thicklines \path(220,113)(240,113)
\thicklines \path(1436,113)(1416,113)
\put(198,113){\makebox(0,0)[r]{-0.3}}
\thicklines \path(220,240)(240,240)
\thicklines \path(1436,240)(1416,240)
\put(198,240){\makebox(0,0)[r]{-0.2}}
\thicklines \path(220,368)(240,368)
\thicklines \path(1436,368)(1416,368)
\put(198,368){\makebox(0,0)[r]{-0.1}}
\thicklines \path(220,495)(240,495)
\thicklines \path(1436,495)(1416,495)
\put(198,495){\makebox(0,0)[r]{0}}
\thicklines \path(220,622)(240,622)
\thicklines \path(1436,622)(1416,622)
\put(198,622){\makebox(0,0)[r]{0.1}}
\thicklines \path(220,750)(240,750)
\thicklines \path(1436,750)(1416,750)
\put(198,750){\makebox(0,0)[r]{0.2}}
\thicklines \path(220,877)(240,877)
\thicklines \path(1436,877)(1416,877)
\put(198,877){\makebox(0,0)[r]{0.3}}
\thicklines \path(220,113)(220,133)
\thicklines \path(220,877)(220,857)
\put(220,68){\makebox(0,0){-20}}
\thicklines \path(372,113)(372,133)
\thicklines \path(372,877)(372,857)
\put(372,68){\makebox(0,0){-15}}
\thicklines \path(524,113)(524,133)
\thicklines \path(524,877)(524,857)
\put(524,68){\makebox(0,0){-10}}
\thicklines \path(676,113)(676,133)
\thicklines \path(676,877)(676,857)
\put(676,68){\makebox(0,0){-5}}
\thicklines \path(828,113)(828,133)
\thicklines \path(828,877)(828,857)
\put(828,68){\makebox(0,0){0}}
\thicklines \path(980,113)(980,133)
\thicklines \path(980,877)(980,857)
\put(980,68){\makebox(0,0){5}}
\thicklines \path(1132,113)(1132,133)
\thicklines \path(1132,877)(1132,857)
\put(1132,68){\makebox(0,0){10}}
\thicklines \path(1284,113)(1284,133)
\thicklines \path(1284,877)(1284,857)
\put(1284,68){\makebox(0,0){15}}
\thicklines \path(1436,113)(1436,133)
\thicklines \path(1436,877)(1436,857)
\put(1436,68){\makebox(0,0){20}}
\thicklines \path(220,113)(1436,113)(1436,877)(220,877)(220,113)
\put(45,495){\makebox(0,0)[l]{\shortstack{$ \bar{\cal A} $}}}
\put(828,23){\makebox(0,0){$ C_X $}}
\Thicklines \path(220,805)(220,805)(271,810)(296,812)(321,815)(334,816)(340,816)(347,816)(353,817)(356,817)(359,817)(362,817)(364,817)(366,817)(367,817)(369,817)(370,817)(372,817)(374,817)(375,817)(377,817)(378,817)(380,817)(381,817)(385,817)(388,817)(391,816)(397,816)(404,815)(410,814)(416,813)(423,811)(429,809)(435,806)(448,799)(454,795)(461,789)(473,775)(486,755)(499,728)(511,691)(524,642)(549,509)(562,431)(575,357)(581,325)(587,298)(594,277)(597,269)(600,263)(603,258)
\Thicklines \path(603,258)(605,256)(606,255)(608,254)(609,253)(611,253)(613,253)(614,253)(616,254)(617,255)(619,256)(622,259)(625,263)(632,274)(638,287)(651,317)(676,382)(701,439)(727,485)(752,521)(777,550)(803,573)(828,591)(879,619)(929,639)(980,653)(1031,665)(1081,673)(1132,680)(1183,686)(1233,691)(1284,695)(1335,699)(1385,702)(1436,705)
\thicklines \dashline[-10]{25}(220,256)(220,256)(271,261)(321,268)(372,277)(423,289)(473,304)(524,325)(575,352)(625,388)(676,434)(727,488)(777,544)(803,570)(828,591)(841,600)(853,607)(866,613)(872,615)(879,616)(885,618)(888,618)(890,618)(891,618)(893,618)(894,618)(896,618)(898,619)(899,619)(901,618)(902,618)(904,618)(907,618)(909,618)(910,618)(917,617)(923,615)(929,614)(942,609)(955,603)(980,588)(1031,551)(1081,512)(1132,476)(1183,444)(1233,418)(1284,396)(1335,378)(1385,363)(1436,351)
\thinlines \dashline[-10]{25}(220,300)(220,300)(271,308)(321,318)(372,330)(423,345)(473,364)(524,388)(575,418)(625,454)(676,496)(727,539)(752,558)(777,574)(790,581)(803,586)(809,588)(815,589)(818,590)(822,590)(825,591)(828,591)(830,591)(831,592)(833,592)(834,592)(836,592)(837,592)(839,592)(841,592)(842,592)(844,592)(847,591)(850,591)(853,591)(860,589)(866,588)(879,583)(891,578)(904,571)(929,553)(980,512)(1031,468)(1081,429)(1132,396)(1183,370)(1233,349)(1284,332)(1335,319)(1385,308)(1436,299)
\thinlines \path(220,726)(220,726)(271,723)(321,719)(372,714)(423,707)(473,699)(524,688)(575,674)(625,657)(676,638)(727,617)(752,608)(777,600)(790,597)(803,594)(809,593)(815,592)(822,592)(825,592)(828,591)(831,591)(833,591)(834,591)(836,591)(837,591)(839,591)(841,591)(842,591)(844,591)(845,591)(847,591)(849,591)(850,591)(853,591)(856,591)(860,591)(866,592)(872,593)(879,594)(904,599)(929,606)(980,623)(1031,641)(1081,657)(1132,671)(1183,682)(1233,692)(1284,699)(1335,705)(1385,710)
\thinlines \path(1385,710)(1436,714)
\thinlines \dashline[-10]{5}(220,541)(220,541)(271,546)(321,550)(372,555)(423,560)(473,566)(524,571)(575,576)(625,581)(676,585)(701,587)(727,589)(752,590)(777,591)(790,591)(796,591)(803,591)(809,591)(812,591)(815,591)(818,591)(820,591)(822,591)(823,591)(825,591)(826,591)(828,591)(830,591)(831,591)(833,591)(834,591)(837,591)(841,591)(844,591)(847,591)(853,591)(860,591)(866,591)(879,591)(904,590)(929,589)(980,585)(1031,581)(1081,576)(1132,571)(1183,566)(1233,560)(1284,555)(1335,550)(1385,546)
\thinlines \dashline[-10]{5}(1385,546)(1436,541)
\thinlines \dashline[-10]{5}(220,541)(220,541)(271,546)(321,550)(372,555)(423,560)(473,566)(524,571)(575,576)(625,581)(676,585)(701,587)(727,589)(752,590)(777,591)(790,591)(796,591)(803,591)(809,591)(812,591)(815,591)(818,591)(820,591)(822,591)(823,591)(825,591)(826,591)(828,591)(830,591)(831,591)(833,591)(834,591)(837,591)(841,591)(844,591)(847,591)(853,591)(860,591)(866,591)(879,591)(904,590)(929,589)(980,585)(1031,581)(1081,576)(1132,571)(1183,566)(1233,560)(1284,555)(1335,550)(1385,546)
\thinlines \dashline[-10]{5}(1385,546)(1436,541)
\thinlines \dashline[-10]{5}(220,541)(220,541)(271,546)(321,550)(372,555)(423,560)(473,566)(524,571)(575,576)(625,581)(676,585)(701,587)(727,589)(752,590)(777,591)(790,591)(796,591)(803,591)(809,591)(812,591)(815,591)(818,591)(820,591)(822,591)(823,591)(825,591)(826,591)(828,591)(830,591)(831,591)(833,591)(834,591)(837,591)(841,591)(844,591)(847,591)(853,591)(860,591)(866,591)(879,591)(904,590)(929,589)(980,585)(1031,581)(1081,576)(1132,571)(1183,566)(1233,560)(1284,555)(1335,550)(1385,546)
\thinlines \dashline[-10]{5}(1385,546)(1436,541)
\thinlines \dashline[-10]{5}(220,541)(220,541)(271,546)(321,550)(372,555)(423,560)(473,566)(524,571)(575,576)(625,581)(676,585)(701,587)(727,589)(752,590)(777,591)(790,591)(796,591)(803,591)(809,591)(812,591)(815,591)(818,591)(820,591)(822,591)(823,591)(825,591)(826,591)(828,591)(830,591)(831,591)(833,591)(834,591)(837,591)(841,591)(844,591)(847,591)(853,591)(860,591)(866,591)(879,591)(904,590)(929,589)(980,585)(1031,581)(1081,576)(1132,571)(1183,566)(1233,560)(1284,555)(1335,550)(1385,546)
\thinlines \dashline[-10]{5}(1385,546)(1436,541)
\end{picture} \\
{Figure 21  ~\small The contribution to the integrated normalized asymmetry
$\bar{ {\cal A} }$ from $C_{LL}$ (thick
solid line), $C_{LR}$ (thick dashed line), $C_{RL}$ (thin
dashed line), $C_{RR}$ (thin solid line) and the tensor
and scalar interactions  (dotted line). }


\section{Summary}

We have given the most general model-independent analysis of the rare
$B$ decay process $\Bsll$. 
This process is experimentally very clean, 
and possibly most sensitive to 
the various extensions of the Standard Model (SM), 
compared to other rare decay processes.
As is well known, the main reason for studying rare $B$ decays is
to measure the effective FCNC vertices in order to test the SM precisely
and to search for new physics beyond the SM.
We have investigated the decay $\Bsll$ in the full operator basis, instead
of doing a model by model analysis.
The sensitivity to the coefficients of ten independent local 
four-Fermi interactions is systematically studied for the branching
ratios and the forward-backward (FB) asymmetries. The correlation between 
the branching ratio and the  FB asymmetry has been studied, and the 
flow for $C_{LL}$, the coefficient of the operator  
$(\bar{s}_L \gamma_\mu b_L ~\bar{l}_L \gamma^\mu l_L)$, 
is found to be distinctive from the other vector type interactions. 
The reason for the strong
sensitivity to $C_{LL}$ comes from the large interference
between $(C_9-C_{10})$ and $C_{LL}$.  The tensor and scalar 
operators increase the breaching ratio, while the asymmetry
is not changed at all. 
The other flows have also been discussed in detail.
 
As further comments, the uncertainties from higher-order QCD
corrections can be reduced and such a reduction is very useful to 
distinguish the effect of new physics from the uncertainties 
in the predictions of the SM. 
Unless the theoretical predictions for the rates 
and the asymmetries in the SM are on firm ground, it would be difficult
to make sure that new physics has indeed been discovered. 
A complete short distance NLO calculation for $\Bsll$ is
available \cite{Misiak,Buras}, 
and the $1/m_b^2$ corrections have also been known
\cite{ATM,Falk} for some time, 
but  these are found to give only very small corrections.
The remaining uncertainties are the non-perturbative long distance
contributions associated  with the $J/\psi$ and $\psi'$ resonances \cite{Long},
where some modeling uncertainties remain, 
and the related $1/m_c^2$ corrections. 

The other useful measurements, such
as the tau polarization asymmetry, CP-violating rate asymmetry \cite{Kruger} 
and triple momentum-spin correlations
can be studied and independent information on
the coefficients can also be obtained \cite{future}. The analysis based on
a specific model may be systematically organized \cite{future} in our
scheme, because the comparison of the predictions of the
various models will be easy in our framework.

\section*{\bf Acknowledgments}

We would like to thank K. Kiers, Y. Okada and A. I. Sanda for
a careful reading of the manuscript and for suggestions.
The work of C.S.K. was supported 
in part by Non-Directed-Research-Fund of year 1997, KRF,
in part by the CTP, Seoul National University, 
in part by the BSRI Program, Ministry of Education, BSRI-98-2425
and in part by the KOSEF-DFG large collaboration project, 
Project No. 96-0702-01-01-2.
The work of T.M. was supported by Grant-in-Aid for
Scientific Research on Priority Areas (Physics of CP violation).
The work of T.Y. was supported in part by Grant-in-Aid 
for Scientific Research from the Ministry of Education, 
Science and Culture of Japan and in part by JSPS Research 
Fellowships for Young Scientists. 
\newpage
\appendix
\section{Appendix}
\renewcommand{\theequation}{\Alph{section}-\arabic{equation}}
\setcounter{equation}{0} 

In this Appendix, we derive the various distributions for
the massive lepton case. The formulae can be applied to
the process
 $ B\rightarrow X_s \tau^+ \tau^- $.\\ 
The double differential rate is a function of the dilepton invariant
mass $s$ and $z=cos\theta $, 
\bea
\frac{d {\cal B}}{d s d z} = \frac{1}{2 {m_b}^8}{\cal B}_0 &[& 
                 A_1(s,z)~\{ 4 \left|C_7\right|^2 \} \nn \\
             &+& A_2(s,z)~\{ \left|(C_9^{eff} - C_{10})\right|^2 
                         +  \left|(C_9^{eff} + C_{10})\right|^2 \} \nn \\
             &+& A_3(s,z)~\{ \left|(C_9^{eff} - C_{10})\right|^2 
                         -  \left|(C_9^{eff} + C_{10})\right|^2 \} \nn \\
             &+& A_4(s,z)~\{ 2 Re(-2 C_7 ( C_9^{eff *} - C_{10}^* )) 
                          + 2 Re(-2 C_7 ( C_9^{eff *} + C_{10}^* )) \} \nn \\
             &+& A_5(s,z)~\{ 2 Re(-2 C_7 ( C_9^{eff *} - C_{10}^* )) 
                          - 2 Re(-2 C_7 ( C_9^{eff *} + C_{10}^* )) 
                                                             \} \nn \\
             &+& L_1(s,z)~\{ 2 Re( ( C_9^{eff } - C_{10} )
                                  ( C_9^{eff *} + C_{10}^* ))
                                                             \} \nn \\[5mm]
             &+& A_2(s,z)~\{ 2 Re(( C_9^{eff} - C_{10}){\bf C_{LL}^* })  
                          + 2 Re(( C_9^{eff} + C_{10}){\bf C_{LR}^* }) 
                                                            \} \nn \\
             &+& A_3(s,z)~\{ 2 Re(( C_9^{eff} - C_{10}){\bf C_{LL}^* })  
                          - 2 Re(( C_9^{eff} + C_{10}){\bf C_{LR}^* })
                                                            \} \nn \\
             &+& A_4(s,z)~\{ 2 Re( -2 C_7({\bf C_{LL}^* + C_{LR}^* }))\} \nn \\
             &+& A_5(s,z)~\{ 2  Re( -2 C_7({\bf C_{LL}^* - C_{LR}^* }))
                                                            \} \nn \\ 
             &+& A_6(s,z)~\{ 2 Re( -2 C_7({\bf C_{RL}^* + C_{RR}^* }))\}
                                                               \nn \\
             &+& A_6(s,z)~\{ - 2 Re(( C_9^{eff} - C_{10}){\bf C_{RL}^* })  
                            - 2 Re(( C_9^{eff} + C_{10}){\bf C_{RR}^* })
                                                            \} \nn \\
             &+& A_7(s,z)~\{ 2 Re( -2 C_7({\bf C_{RL}^* - C_{RR}^* })) 
                                                         \} \nn \\
             &+& L_2(s,z)~\{ 4m_b Re( -2 C_7({\bf C_{LRLR}^* + C_{LRRL}^*
                                                          }) \} \nn \\
             &+& L_2(s,z)~\{ 4 m_s Re( -2 C_7({\bf C_{RLLR}^* + C_{RLRL}^*
                                                          }) \} \nn \\ 
             &+& L_3(s,z)~\{ 2 Re( -2 C_7({\bf C_{T}^* 
                                                          }) \} \nn \\
             &+& L_4(s,z)~\{ 2 Re( -2 C_7({\bf C_{TE}^*
                                                          }) \} \nn \\ 
             &+& L_1(s,z)~\{ 2 Re(( C_9^{eff} - C_{10}){\bf C_{LR}^* })  
                           + 2 Re(( C_9^{eff} + C_{10}){\bf C_{LL}^* })
                                                            \} \nn \\
             &+& L_5(s,z)~\{ - 2 Re(( C_9^{eff} - C_{10}){\bf C_{RL}^* })  
                            - 2 Re(( C_9^{eff} + C_{10}){\bf C_{RR}^* })
                                                            \} \nn \\
             &+& L_5(s,z)~\{ 2 Re(( C_9^{eff} - C_{10}){\bf C_{RR}^* })  
                           + 2 Re(( C_9^{eff} + C_{10}){\bf C_{RL}^* })
                                                            \} \nn \\ 
             &+& L_6(s,z)~\{ 2 Re(( C_9^{eff} - C_{10})
                                  ({\bf C_{LRLR}^* - C_{LRRL}^*}) \nn \\
             & & ~~~~~~~~~ - 2 Re(( C_9^{eff} + C_{10})
                                  ({\bf C_{LRLR}^* - C_{LRRL}^*})
                                                            \} \nn \\ 
              &+& L_2(s,z)~(- m_b)\{ 2 Re(( C_9^{eff} - C_{10})
                                  ({\bf C_{LRLR}^* +  C_{LRRL}^*}) \nn 
                                      \\
             & & ~~~~~~~~~~~~~~~ + 2 Re(( C_9^{eff} + C_{10})
                                  ({\bf C_{LRLR}^* + C_{LRRL}^*})
                                                            \} \nn \\ 
             &+& L_7(s,z)~\{ 2 Re(( C_9^{eff} - C_{10})
                                  ({\bf C_{RLRL}^* - C_{RLLR}^*}) \nn \\
             & & ~~~~~~~~~ - 2 Re(( C_9^{eff} + C_{10})
                                  ({\bf C_{RLRL}^* - C_{RLLR}^*})
                                                            \} \nn \\ 
              &+& L_2(s,z)~(- m_s)\{ 2 Re(( C_9^{eff} - C_{10})
                                  ({\bf C_{RLLR}^* +  C_{RLRL}^*}) \nn 
                                  \\
             & & ~~~~~~~~~~~~~~~ + 2 Re(( C_9^{eff} + C_{10})
                                  ({\bf C_{RLLR}^* + C_{RLRL}^*})
                                                            \} \nn \\ 
             &+& L_8(s,z)~\{ 2 Re(( C_9^{eff} - C_{10})
                                  {\bf C_{T}^* })  
                           + 2 Re(( C_9^{eff} + C_{10})
                                  {\bf C_{T}^* })
                                                            \} \nn \\ 
              &+& L_2(s,z)~ 12 (m_b + m_s)\{ 2 Re(( C_9^{eff} - C_{10})
                                  {\bf C_{T}^* })  
                           - 2 Re(( C_9^{eff} + C_{10})
                                  {\bf C_{T}^* })
                                                            \} \nn \\ 
             &+& L_9(s,z)~\{ 2 Re(( C_9^{eff} - C_{10})
                                  {\bf C_{TE}^* })  
                           + 2 Re(( C_9^{eff} + C_{10})
                                  {\bf C_{TE}^* })
                                                            \} \nn \\ 
              &+& L_2(s,z)~24(-m_b+m_s)\{ 2 Re(( C_9^{eff} - C_{10})
                                  {\bf C_{TE}^* })  
                           - 2 Re(( C_9^{eff} + C_{10})
                                  {\bf C_{TE}^*})
                                                            \} \nn \\[5mm]
             &+& A_2(s,z)~\{ {\bf \left|C_{LL}\right|^2 
             + \left|C_{LR}\right|^2 
             + \left|C_{RL}\right|^2 + \left|C_{RR}\right|^2  } \} \nn \\
             &+& A_3(s,z)~\{ {\bf  \left|C_{LL}\right|^2 
                                - \left|C_{LR}\right|^2 
                                -  \left|C_{RL}\right|^2 
                                + \left|C_{RR}\right|^2 } \} \nn \\
             &+& A_6(s,z)~\{ - 2 {\bf Re( C_{LL} C_{RL}^* 
                                      + C_{LR} C_{RR}^* )} \nn \\
             & & ~~~~~~~~~~ + {\bf Re(C_{LRLR} C_{RLLR}^* 
                            + C_{LRRL} C_{RLRL}^* ) } \} \nn \\
             &+& A_8(s,z)~\{ {\bf \left|C_{LRLR}\right|^2 
                                + \left|C_{RLLR}\right|^2 
                                + \left|C_{LRRL}\right|^2        
                                + \left|C_{RLRL}\right|^2  } \} \nn \\
              &+& A_3(s,z)~\{ - 4 {\bf Re( C_{LRLR})(
                                   C_{T}^* - 2 C_{TE}^* )} 
                           - 4 {\bf Re(C_{RLRL})(
                                 C_{T} +2 C_{TE}^* ) } \} \nn \\    
             &+& A_9(s,z)~\{ 16 {\bf \left|C_{T}\right|^2 
                                } \} \nn \\
             &+& A_9(s,z)~\{ 64 {\bf 
                                \left|C_{TE}\right|^2 } \} \nn \\
             &+& L_1(s,z)~\{ 2 {\bf Re( C_{LL} C_{LR}^* 
                                      + C_{RL} C_{RR}^* )} \nn \\
             & & ~~~~~~~~~~ - {\bf Re(C_{LRLR} C_{LRRL}^* 
                            + C_{RLLR} C_{RLRL}^* ) } \} \nn \\
             &+& L_5(s,z)~\{ - 2 {\bf Re( C_{LL} C_{RL}^* 
                                      + C_{LR} C_{RR}^* )}\nn \\ 
             & & ~~~~~~~~~~~~~ + {\bf Re(C_{LRLR} C_{RLLR}^* 
                            + C_{LRRL} C_{RLRL}^* ) } \} \nn \\
             &+& L_5(s,z)~\{ 2 {\bf Re( C_{LL} C_{RR}^* 
                                      + C_{LR} C_{RL}^* )} \nn \\
             & & ~~~~~~~~~~~ + \frac{1}{2}{\bf Re(C_{LRLR} C_{RLRL}^* 
                            + C_{LRRL} C_{RLLR}^* ) } \} \nn \\
             &+& L_6(s,z)~\{ 2 {\bf Re( C_{LL} - C_{LR})(
                                      C_{LRLR}^* - C_{LRRL}^* )} \nn \\
             & & ~~~~~~~~~~ + 2 {\bf Re(C_{RL} - C_{RR})(
                                 C_{RLLR}^* - C_{RLRL}^* ) } \} \nn \\
             &+& L_2(s,z)~(- m_b)\{ 2 {\bf Re( C_{LL} + C_{LR})(
                                      C_{LRLR}^* + C_{LRRL}^* )} \nn \\
             & & ~~~~~~~~~~~~~~~~~ + 2 {\bf Re(C_{RL} + C_{RR})(
                                 C_{RLLR}^* + C_{RLRL}^* ) } \} \nn \\
             &+& L_7(s,z)~\{ 2 {\bf Re( C_{LL} - C_{LR})(
                                      C_{RLRL}^* + C_{RLLR}^* )} \nn \\
             & & ~~~~~~~~~~~ + 2 {\bf Re(C_{RL} - C_{RR})(
                                 C_{LRRL}^* - C_{LRLR}^* ) } \} \nn \\
             &+& L_2(s,z)~(- m_s)\{ 2 {\bf Re( C_{LL} + C_{LR})(
                                      C_{RLRL}^* + C_{RLLR}^* )} \nn \\
             & & ~~~~~~~~~~~~~~~~~ + 2 {\bf Re(C_{RL} + C_{RR})(
                                 C_{LRRL}^* - C_{LRLR}^* ) } \} \nn \\
             &+& L_8(s,z)~\{ 2 {\bf Re( C_{LL} + C_{LR})(
                                      C_{T}^* )} 
                           + 2 {\bf Re(C_{RL} + C_{RR})(
                                 C_{T}^* ) } \} \nn \\
             &+& L_2(s,z)~12(m_b+m_s)\{ 2 {\bf Re( C_{LL} - C_{LR})(
                                      C_{T}^* )} 
                            + 2 {\bf Re(C_{RL} - C_{RR})(
                                 C_{T}^* ) } \} \nn \\       
              &+& L_9(s,z)~\{ 2 {\bf Re( C_{LL} + C_{LR})(
                                      C_{TE}^* )} 
                           + 2 {\bf Re(C_{RL} + C_{RR})(
                                 C_{TE}^* ) } \} \nn \\
             &+& L_2(s,z)~24(-m_b+m_s)\{ 2 {\bf Re( C_{LL} - C_{LR})(
                                      C_{TE}^* )} \nn \\
             & & ~~~~~~~~~~~~~ + 2 {\bf Re(C_{RL} - C_{RR})(
                                 C_{TE}^* ) } \} \nn \\ 
             &+& L_5(s,z)~\{ {\bf - 192 
                               \left|C_{TE}\right|^2 }  \} ],
\eea
where $A_n$ and $L_n$ are functions of the kinematic variables $s$ and $z$,
\bea
u(s) &=& \sqrt{( s - (m_b + m_s )^2 )( s - (m_b - m_s )^2 )
               ( 1 - \frac{4 m_l^2 }{s} )} , \nn \\
A_1 &=& \frac{1}{s^2}u(s)[ - 16 m_b^2 m_s^2 s ( 2 m_l^2 + s ) \nn \\
   & & + ( m_b^2 + m_s^2 ) \{ m_l^2 ( - 8 ( m_b^2 + m_s^2 ) s + 
                                   8 ( m_b^2 - m_s^2 )^2 ) 
       + 2 s ( - s^2 + u(s)^2 z^2 + ( m_b^2 - m_s^2)^2 )\}] ,  \nn \\
A_2 &=& u(s) ( - u(s)^2 z^2 - s^2 + ( m_b^2 - m_s^2)^2 ) , \nn \\
A_3 &=&  2 u(s)^2 z s  , \nn \\
A_4 &=& u(s) \frac{1}{s}\{ 2 (2 m_l^2 + s )
                  ( ( m_b^2 + m_s^2 ) s - ( m_b^2 - m_s^2 )^2 ) , 
                          \nn \\
A_5 &=& -2 ( m_b^2 + m_s^2 ) u(s)^2 z ,  \nn \\
A_6 &=& 4 u(s) m_b m_s (2m_l^2 + s ) ,  \nn \\
A_7 &=& 4 m_b m_s u(s)^2 z  , \nn \\
A_8 &=& - u(s) ( m_b^2 + m_s^2 -s )( 2 m_l^2 - s )  , \nn \\
A_9 &=& u(s) \{ 4 m_l^2( m_b^2 - 6 m_b m_s + m_s^2 - s ) \nn \\
  & & - 2 u(s)^2 z^2 - 2 ( m_b^2 + m_s^2) s 
      + 2  ( m_b^2 - m_s^2 )^2 \}  , \nn \\
L_1 &=& 4 u(s) m_l^2 ( m_b^2 + m_s^2 -s )  , \nn \\
L_2 &=& m_l u(s)^2 z  , \nn \\
L_3 &=& u(s) \frac{1}{s} m_l ( m_b + m_s)( 8 s^2 + 8 s 
            ( m_b^2 + 6 m_b m_s + m_s^2 ) 
         - 16 ( m_b^2 - m_s^2)^2 )  , \nn \\
L_4 &=& u(s) \frac{1}{s} m_l ( m_b - m_s)( - 16 s^2 - 16 s 
            ( m_b^2 - 6 m_b m_s + m_s^2 ) 
         + 32 ( m_b^2 - m_s^2)^2
                               )  , \nn \\
L_5 &=& -16 u(s) m_b m_s m_l^2  , \nn \\
L_6 &=& u(s) m_l m_b ( s - m_b^2 + m_s^2 )  , \nn \\
L_7 &=& u(s) m_l m_s ( s + m_b^2 - m_s^2 )  , \nn \\
L_8 &=& 12 u(s) m_l ( - ( m_b + m_s ) s + (
                           m_b -  m_s ) (m_b^2 -  m_s^2 )  , \nn \\
L_9 &=& 24 u(s) m_l ( ( m_b - m_s ) s - (
                           m_b + m_s ) (m_b^2 -  m_s^2 )  . 
\eea
Among them, $A_1,A_2,A_4,A_6,A_8,A_9,L_1,L_3,L_4,L_5,L_6,L_7,L_8$ and $L_9$ are even
functions with respect to $z$, while $A_3,A_5,A_7,L_2$ are odd functions.
In the limit of $m_l=0$,  $L_n$ vanish and $A_n$ lead to those in the
massless limit.

To investigate the differential decay rate and the FB asymmetry, 
we define the following functions.
They are obtained by integrateing the $A_n(s,z)$ ($n$ is 1 to 9) and $L_n(s,z)$
with respect to z as follows,
\bea
M_i(s) &=& \int_{-1}^1 A_i(s,z) dz = 2 A_i(s,\sqrt{1/3}) , \nn \\
N_j(s) &=& \int_{-1}^1 L_j(s,z) dz = 2 L_j(s,\sqrt{1/3}) ,
\eea 
where $i$ is 1,2,4,6,8 and 9, and $j$ is 1,3,4,5,6,7,8 and 9. Also
\bea
A_k(s,1) &=&  \int_{0}^1 A_k(s,z) dz - \int_{-1}^0 A_k(s,z) dz , \nn \\
L_2(s,1) &=&  \int_{0}^1 L_2(s,z) dz - \int_{-1}^0 L_2(s,z) dz,
\eea
where $k$ is 3,5 and 7.

Finally, the differential decay rate is given by 
\bea
\frac{d {\cal B}}{d s } = \frac{1}{2{m_b}^8}{\cal B}_0 &[& 
                 M_1(s)~\{ 4 \left|C_7\right|^2 \} \nn \\
             &+& M_2(s)~\{ \left|(C_9^{eff} - C_{10})\right|^2 
                         +  \left|(C_9^{eff} + C_{10})\right|^2 \} \nn \\
             &+& M_4(s)~\{ 2 Re(-2 C_7 ( C_9^{eff *} - C_{10}^* )) 
                          + 2 Re(-2 C_7 ( C_9^{eff *} + C_{10}^* )) \} \nn \\
             &+& N_1(s)~\{ 2 Re( ( C_9^{eff } - C_{10} )
                                  ( C_9^{eff *} + C_{10}^* ))
                                                             \} \nn \\[5mm]
             &+& M_2(s)~\{ 2 Re(( C_9^{eff} - C_{10}){\bf C_{LL}^* })  
                          + 2 Re(( C_9^{eff} + C_{10}){\bf C_{LR}^* }) 
                                                            \} \nn \\
             &+& M_4(s)~\{ 2 Re( -2 C_7({\bf C_{LL}^* + C_{LR}^* }))\} \nn \\
             &+& M_6(s)~\{ 2 Re( -2 C_7({\bf C_{RL}^* + C_{RR}^* }))\}
                                                               \nn \\
             &+& M_6(s)~\{ - 2 Re(( C_9^{eff} - C_{10}){\bf C_{RL}^* })  
                            - 2 Re(( C_9^{eff} + C_{10}){\bf C_{RR}^* })
                                                            \} \nn \\
             &+& N_3(s)~\{ 2 Re( -2 C_7({\bf C_{T}^* 
                                                          }) \} \nn \\
             &+& N_4(s)~\{ 2 Re( -2 C_7({\bf C_{TE}^*
                                                          }) \} \nn \\ 
             &+& N_1(s)~\{ 2 Re(( C_9^{eff} - C_{10}){\bf C_{LR}^* })  
                           + 2 Re(( C_9^{eff} + C_{10}){\bf C_{LL}^* })
                                                            \} \nn \\
             &+& N_5(s)~\{ - 2 Re(( C_9^{eff} - C_{10}){\bf C_{RL}^* })  
                            - 2 Re(( C_9^{eff} + C_{10}){\bf C_{RR}^* })
                                                            \} \nn \\
             &+& N_5(s)~\{ 2 Re(( C_9^{eff} - C_{10}){\bf C_{RR}^* })  
                           + 2 Re(( C_9^{eff} + C_{10}){\bf C_{RL}^* })
                                                            \} \nn \\ 
             &+& N_6(s)~\{ 2 Re(( C_9^{eff} - C_{10})
                                  ({\bf C_{LRLR}^* - C_{LRRL}^*}) \nn \\ 
             & & ~~~~~~~ - 2 Re(( C_9^{eff} + C_{10})
                                  ({\bf C_{LRLR}^* - C_{LRRL}^*})
                                                            \} \nn \\ 
             &+& N_7(s)~\{ 2 Re(( C_9^{eff} - C_{10})
                                  ({\bf C_{RLRL}^* - C_{RLLR}^*}) \nn \\  
             & & ~~~~~~~ - 2 Re(( C_9^{eff} + C_{10})
                                  ({\bf C_{RLRL}^* - C_{RLLR}^*})
                                                            \} \nn \\ 
             &+& N_8(s)~\{ 2 Re(( C_9^{eff} - C_{10})
                                  {\bf C_{T}^* })  
                           + 2 Re(( C_9^{eff} + C_{10})
                                  {\bf C_{T}^* })
                                                            \} \nn \\ 
             &+& N_9(s)~\{ 2 Re(( C_9^{eff} - C_{10})
                                  {\bf C_{TE}^* })  
                           + 2 Re(( C_9^{eff} + C_{10})
                                  {\bf C_{TE}^* })
                                                            \} \nn \\[5mm] 
             &+& M_2(s)~\{ {\bf \left|C_{LL}\right|^2 + \left|C_{LR}\right|^2 
             + \left|C_{RL}\right|^2 + \left|C_{RR}\right|^2  } \} \nn \\
             &+& M_6(s)~\{ - 2 {\bf Re( C_{LL} C_{RL}^* 
                                      + C_{LR} C_{RR}^* )} \nn \\
             & & ~~~~~~~~~ + {\bf Re(C_{LRLR} C_{RLLR}^* 
                            + C_{LRRL} C_{RLRL}^* ) } \} \nn \\
             &+& M_8(s)~\{ {\bf \left|C_{LRLR}\right|^2 
                                + \left|C_{RLLR}\right|^2 
                                + \left|C_{LRRL}\right|^2        
                                + \left|C_{RLRL}\right|^2  } \} \nn \\
             &+& M_9(s)~\{ 16 {\bf \left|C_{T}\right|^2 
                                } \} \nn \\
             &+& M_9(s)~\{ 64 {\bf 
                                \left|C_{TE}\right|^2 } \} \nn \\
             &+& N_1(s)~\{ 2 {\bf Re( C_{LL} C_{LR}^* 
                                      + C_{RL} C_{RR}^* )} \nn \\
             & & ~~~~~~~~ - {\bf Re(C_{LRLR} C_{LRRL}^* 
                            + C_{RLLR} C_{RLRL}^* ) } \} \nn \\
             &+& N_5(s)~\{ - 2 {\bf Re( C_{LL} C_{RL}^* 
                                      + C_{LR} C_{RR}^* )} \nn \\
             & & ~~~~~~~~~~~ + {\bf Re(C_{LRLR} C_{RLLR}^* 
                            + C_{LRRL} C_{RLRL}^* ) } \} \nn \\
             &+& N_5(s)~\{ 2 {\bf Re( C_{LL} C_{RR}^* 
                                      + C_{LR} C_{RL}^* )} \nn \\
             & & ~~~~~~ + \frac{1}{2}{\bf Re(C_{LRLR} C_{RLRL}^* 
                            + C_{LRRL} C_{RLLR}^* ) } \} \nn \\
             &+& N_6(s)~\{ 2 {\bf Re( C_{LL} - C_{LR})(
                                      C_{LRLR}^* - C_{LRRL}^* )} \nn \\
             & & ~~~~~~ + 2 {\bf Re(C_{RL} - C_{RR})(
                                 C_{RLLR}^* - C_{RLRL}^* ) } \} \nn \\
             &+& N_7(s)~\{ 2 {\bf Re( C_{LL} - C_{LR})(
                                      C_{RLRL}^* + C_{RLLR}^* )} \nn \\
             & & ~~~~~~ + 2 {\bf Re(C_{RL} - C_{RR})(
                                 C_{LRRL}^* - C_{LRLR}^* ) } \} \nn \\
             &+& N_8(s)~\{ 2 {\bf Re( C_{LL} + C_{LR})(
                                      C_{T}^* )} 
                           + 2 {\bf Re(C_{RL} + C_{RR})(
                                 C_{T}^* ) } \} \nn \\
              &+& N_9(s)~\{ 2 {\bf Re( C_{LL} + C_{LR})(
                                      C_{TE}^* )} 
                           + 2 {\bf Re(C_{RL} + C_{RR})(
                                 C_{TE}^* ) } \} \nn \\
             &+& N_5(s)~\{ {\bf - 192 
                               \left|C_{TE}\right|^2 } \} ]. 
\eea

\newpage
\noindent And the FB asymmetry is
\bea
\frac{d {\cal A}}{d s } = \frac{1}{2{m_b}^8}{\cal B}_0 &[& 
                 A_3(s,1)~\{ \left|(C_9^{eff} - C_{10})\right|^2 
                         -  \left|(C_9^{eff} + C_{10})\right|^2 \} \nn \\
             &+& A_5(s,1)~\{ 2 Re(-2 C_7 ( C_9^{eff *} - C_{10}^* )) 
                          - 2 Re(-2 C_7 ( C_9^{eff *} + C_{10}^* )) 
                                                             \} \nn \\[5mm]
             &+& A_3(s,1)~\{ 2 Re(( C_9^{eff} - C_{10}){\bf C_{LL}^* })  
                          - 2 Re(( C_9^{eff} + C_{10}){\bf C_{LR}^* })
                                                            \} \nn \\
             &+& A_5(s,1)~\{ 2  Re( -2 C_7({\bf C_{LL}^* - C_{LR}^* }))
                                                            \} \nn \\ 
             &+& A_7(s,1)~\{ 2 Re( -2 C_7({\bf C_{RL}^* - C_{RR}^* })) 
                                                         \} \nn \\
             &+& L_2(s,1)~\{ 4m_b Re( -2 C_7({\bf C_{LRLR}^* + C_{LRRL}^*
                                                          }) \} \nn \\
             &+& L_2(s,1)~\{ 4 m_s Re( -2 C_7({\bf C_{RLLR}^* + C_{RLRL}^*
                                                          }) \} \nn \\ 
              &+& L_2(s,1)~(-m_b)\{ 2 Re(( C_9^{eff} - C_{10})
                                  ({\bf C_{LRLR}^* +  C_{LRRL}^*}) \nn 
                                  \\ 
              & & ~~~~~~~~~~~~~~~ + 2 Re(( C_9^{eff} + C_{10})
                                  ({\bf C_{LRLR}^* + C_{LRRL}^*})
                                                            \} \nn \\ 
              &+& L_2(s,1)~(-m_s)\{ 2 Re(( C_9^{eff} - C_{10})
                                  ({\bf C_{RLLR}^* +  C_{RLRL}^*}) 
                                       \nn \\ 
              & & ~~~~~~~~~~~~~~~ + 2 Re(( C_9^{eff} + C_{10})
                                  ({\bf C_{RLLR}^* + C_{RLRL}^*})
                                                            \} \nn \\ 
              &+& L_2(s,1)~ 12 (m_b + m_s)\{ 2 Re(( C_9^{eff} - C_{10})
                                  {\bf C_{T}^* })  
                           - 2 Re(( C_9^{eff} + C_{10})
                                  {\bf C_{T}^* })
                                                            \} \nn \\ 
              &+& L_2(s,1)~24(-m_b+m_s)\{ 2 Re(( C_9^{eff} - C_{10})
                                  {\bf C_{TE}^* })  
                           - 2 Re(( C_9^{eff} + C_{10})
                                  {\bf C_{TE}^*})
                                                            \} \nn \\[5mm]
             &+& A_3(s,1)~\{ {\bf  \left|C_{LL}\right|^2 
                                - \left|C_{LR}\right|^2 
                                -  \left|C_{RL}\right|^2 
                                + \left|C_{RR}\right|^2 } \} \nn \\
              &+& A_3(s,1)~\{ - 4 {\bf Re( C_{LRLR})(
                                   C_{T}^* - 2 C_{TE}^* )} 
                           - 4 {\bf Re(C_{RLRL})(
                                 C_{T} +2 C_{TE}^* ) } \} \nn \\    
             &+& L_2(s,1)~(-m_b)\{ 2 {\bf Re( C_{LL} + C_{LR})(
                                      C_{LRLR}^* + C_{LRRL}^* )} \nn \\
             & & ~~~~~~~~~~~~~~~~~ + 2 {\bf Re(C_{RL} + C_{RR})(
                                 C_{RLLR}^* + C_{RLRL}^* ) } \} \nn \\
             &+& L_2(s,1)~(-m_s)\{ 2 {\bf Re( C_{LL} + C_{LR})(
                                      C_{RLLR}^* + C_{RLRL}^* )} \nn \\
             & & ~~~~~~~~~~~~~~~~~ + 2 {\bf Re(C_{RL} + C_{RR})(
                                 C_{LRLR}^* + C_{LRRL}^* ) } \} \nn \\
             &+& L_2(s,1)~12(m_b+m_s)\{ 2 {\bf Re( C_{LL} - C_{LR})(
                                      C_{T}^* )} 
                            + 2 {\bf Re(C_{RL} - C_{RR})(
                                 C_{T}^* ) } \} \nn \\       
             &+& L_2(s,1)~24(-m_b+m_s)\{ 2 {\bf Re( C_{LL} - C_{LR})(
                                      C_{TE}^* )} \nn \\
             & & ~~~~~~~~~~~~~~~~~~~~~~ + 2 {\bf Re(C_{RL} - C_{RR})(
                                 C_{TE}^* ) } \} ].  
\eea



\begin{thebibliography}{99}
\bibitem{Okada}
       The authors would like to thank Y. Okada for
       suggesting to them to draw the correlation.
\end{thebibliography}
\end{document}